\documentclass[aps,prl,superscriptaddress,twocolumn,showpacs, longbibliography]{revtex4-2}

\usepackage{graphicx}
\usepackage{dcolumn}
\usepackage{bm}
\usepackage{amsmath}
\usepackage{longtable}
\newcommand{\KI}{\ensuremath{\,\textrm{KI}}}
\begin{document}

\title{Observation of Quantum Oscillations in The Low Temperature Specific Heat of SmB\textsubscript{6}}

\author{P. G.  LaBarre}
\email{pglabarre@gmail.com}
\affiliation{Department of Physics, University of California at Santa Cruz, Santa Cruz, CA, USA}

\author{A. Rydh}
\affiliation{Department of Physics, Stockholm University, Stockholm, Sweden}

\author{J. Palmer-Fortune}
\affiliation{Department of Physics, Smith College, Northampton MA 01063}

\author{J. A. Frothingham}
\affiliation{Department of Physics, Smith College, Northampton MA 01063}


\author{S. T. Hannahs} 
\affiliation{National High Magnetic Field Laboratory, Florida State University, Tallahassee, FL 32310-3706, USA}



\author{A. P. Ramirez }
\email{apr@ucsc.edu}
\affiliation{Department of Physics, University of California at Santa Cruz, Santa Cruz, CA, USA}

\author{N. Fortune}
\email{nfortune@smith.edu}
\affiliation{Department of Physics, Smith College, Northampton MA 01063}

\date{\today}

\begin{abstract}
We report measurements of the low-temperature specific heat of Al-flux-grown samples of SmB\textsubscript{6} in magnetic fields up to 32 T.  Quantum oscillations periodic in \emph{1/H} are observed between 8 and 32 T at selected angles between [001] and [111].  The observed frequencies and their angular dependence are consistent with previous magnetic torque measurements of SmB\textsubscript{6} but the effective masses inferred from Lifshitz-Kosevich theory are significantly larger and closer to those inferred from zero-field specific heat. Our results are thus consistent with a bulk density of states origin for the oscillations.   \end{abstract}

\maketitle
Despite five decades of study, the Kondo insulator SmB\textsubscript{6} continues to yield new physics, the most recent discovery being the observation of magneto-quantum oscillations (MQOs) characteristic of metals. Curiously, such oscillations are observed in measurements of the magnetic torque \cite{hartstein_intrinsic_2020} but not in charge transport \cite{Wolgast_2017}. 

The physical origin of these oscillations continues to be debated \cite{li_two-dimensional_2014, tan_unconventional_2015,xiang_bulk_2017, hartstein_fermi_2018, thomas_quantum_2019, li_emergent_2020, Knolle_2015, Knolle_2017, Chowdhury_2018}. This is in part because magnetic torque measures the magnetization anisotropy, rather than the bulk susceptibility. Since SmB\textsubscript{6} is a known topological Kondo insulator, it should possess surface or interface states leading to large demagnetization effects in the anisotropy, greatly overestimating the inferred magnetization \cite{li_emergent_2020}. Thus, in addition to a possible bulk origin, one must also consider the possibility that torque-QOs are due to such surface states. Since the geometric properties of these states are not known, torque alone cannot distinguish between bulk or surface, but specific heat QOs are only susceptible to a bulk origin, thus motivating the present study \cite{sullivan_steady-state_1968, shoenberg_magnetic_1984, varma_majorana_2020}.

For a normal metal, Lifshitz-Kosevich (L-K) theory predicts the magnitude of the quantum oscillations in the heat capacity to be on the order of 0.1\% of the ordinary electronic specific heat for good metals \cite{shoenberg_magnetic_1984}. Such oscillations are in principle resolvable using high-resolution ac-calorimetry and Fourier analysis, and have in fact been observed previously in lower carrier density semimetals  \cite{sullivan_steady-state_1968, shoenberg_magnetic_1984} and molecular conductors \cite{fortune_specific-heat_1990, bondarenko_first_2001}.  Our working assumption is that the observed oscillatory behavior in SmB\textsubscript{6} arises from regions of the sample that can support large mean free paths of Fermi liquid-like excitations (even if they are charge neutral in origin) and will therefore still be governed by L-K theory regarding oscillation amplitudes and frequencies even if the material itself is an insulator. 

\begin{figure}[htb!]

\includegraphics[width=\columnwidth]{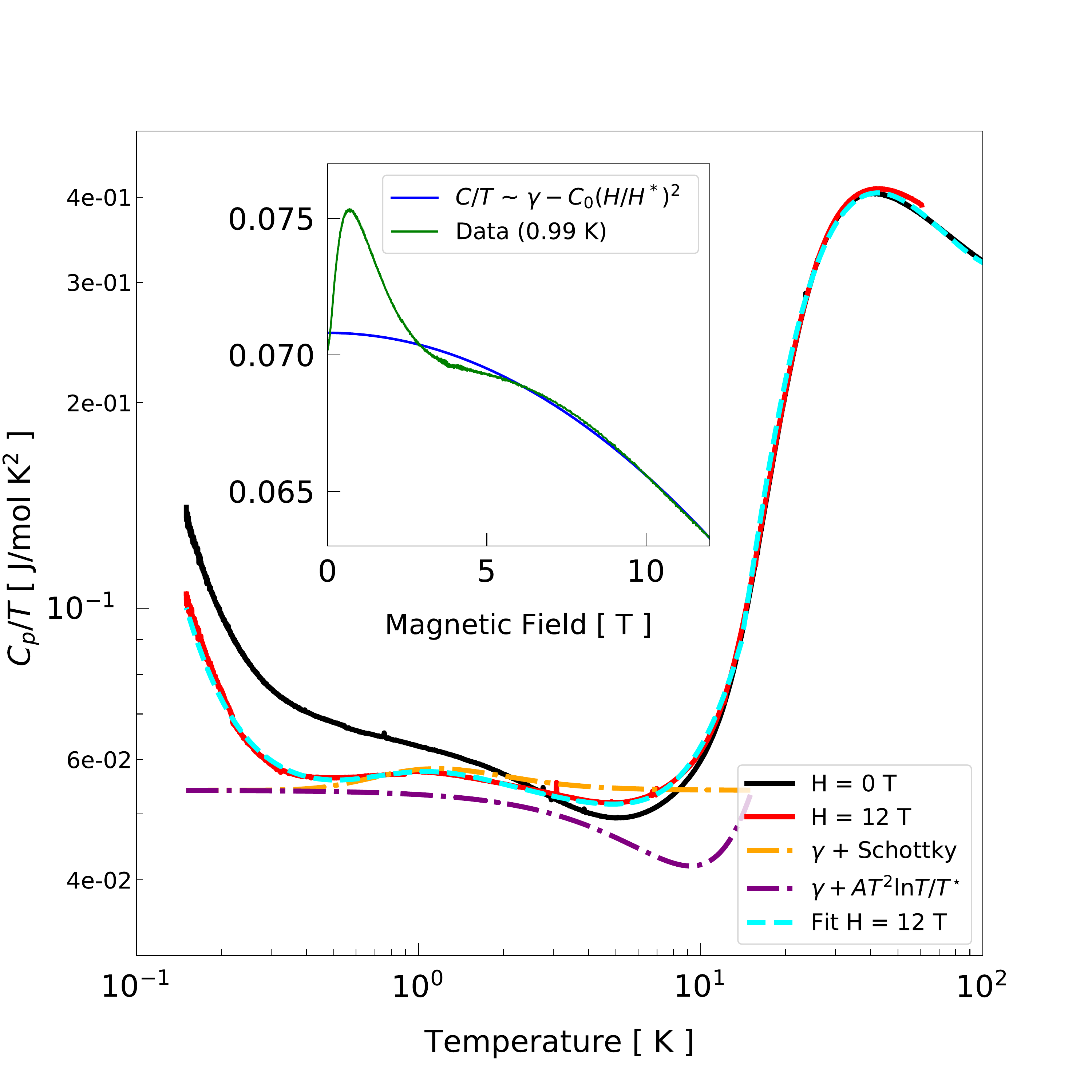}
\caption{\label{fig:zerofield}Temperature dependence of the specific heat at 0~T and 12 T for a 0.430 $\mu$g flux-grown SmB\textsubscript{6} sample, along with a representative fit to the temperature dependence at 12 T. The results are consistent with the presence of spin fluctuations with a characteristic temperature \emph{\(T^{\ast}\)} = 15 K. Inset: field dependence of $C/T$ at $T = 0.99$ K. Above a Schottky peak near 1 T arising from magnetic impurities,  $C/T$ varies as  $\gamma(H) = \gamma(0)(1 - \alpha(H/H^*)^2$, where $H^* = k_BT^*/\mu_B$.  }
\end{figure}

In order to investigate this possibility of bulk oscillations \cite{varma_majorana_2020}, we have measured the specific heat of SmB\textsubscript{6} in applied magnetic fields for  three flux grown crystals of 0.43 $\mu$g, 1.511 $\mu$g, and 0.126 $\mu$g, each grown in a separate batch at Los Alamos National Laboratory \cite{PhysRevB.99.045138}.  Our measurements of $C(T,H)$ at $T < 1K$ and $H$ up to 32 tesla were carried out using custom-built rotatable micro- and nano-calorimeters \cite{tagliati_differential_2012, fortune_top-loading_2014} and also as a function of temperature between 0.1 K and 100 K in various fixed magnetic fields. The heat capacities of the bare calorimeters were measured in separate runs and subtracted from the data; Corrections were also made for the magnetoresistance of the thermometers \cite{fortune_high_2000, fortune_top-loading_2014}.

In Fig.~\ref{fig:zerofield} we compare zero field and 12 T data for H $\parallel$ a axis, along with a representative fit to the 12 T data for a 0.430 $\mu$g sample. We note a large variation in the reported low temperature zero-field electronic specific heat $\gamma$ values  for different samples.  Such a large sample-to-sample variation is unusual for most materials, but is characteristic of  SmB$_6$  and may be due to the variation in number and type of rare-earth impurities or in the density of mid-gap states \cite{Rosa2021BulkAS}.

Our results at both 0 T and 12 T are well described at low \emph{T} by the following model:
\begin{equation}
\label{eq:model_fit}
\begin{split}
C &  =C_{el} + C_{\KI} + \beta_{D}T^3 +  DT^{-2} \\
C_{el} & = \gamma_0 T\ \left[\left(m^{\star }/{m}\right)+A T^2 \ln{\left( {T}/{T^{\star }} \right)} \right] 
\end{split}
\end{equation}
Here $\gamma_0$ in $C_{el}$ is the ``bare'' electronic coefficient of the specific heat expected from band structure, \( m^{\ast}/{m}=\gamma\left(H\right)/{\gamma_0}\) is the many-body effective mass enhancement above the band mass $m$, $A$ is a coupling constant dependent on the strength of the exchange interaction between Fermi-liquid quasiparticles and mass-enhancing excitations, $D/T^2$ is an empirically determined fitting term for the lowest temperature behavior,  
and \emph{\(T^{\ast}\)} is the characteristic temperature for the excitations \cite{ikeda_quenching_1991}. 

The three non-electronic terms in Eq.~\ref{eq:model_fit} include, first,  a highly sample-dependent Schottky-like term $C$\textsubscript{KI} arising from the temperature dependent screening of magnetic impurities in a Kondo insulator \cite{fuhrman_magnetic_2020}. Numerically, the low and high temperature limits of this model closely match the standard Schottky expression for a two-level system with an energy gap~\(\Delta\) and ground state/excited state degeneracy ratio g\textsubscript{0}/g\textsubscript{1} = 2 \cite{gopal_esr_specific_1966}. We have therefore used the Schottky expression as a proxy for this model (which lacks a numerical prediction for intermediate temperatures).   Second, we include a term \(\beta_{D}T^3\) to represent the low temperature limit of the lattice specific heat in the Debye approximation. Third, we add an empirically fit $DT^{-2}$ term to represent an anomalous  upturn in $C$ with decreasing $T$   \cite{FLACHBART_2002,Gabni_2001,flachbart_specific_2006} analogous to but steeper than previously  seen in heavy fermion systems \cite{FLACHBART_2002,Gabni_2001,flachbart_specific_2006, stewart_non-fermi-liquid_2001}. Nuclear Schottky contributions observed at still lower temperatures in applied magnetic fields  \cite{flachbart_specific_2006} have the same $T^{-2}$ dependence but are considered to be too small to be observed here in our data \cite{hartstein_fermi_2018}.

Turning now to the electronic contributions to the specific heat, we note the growing evidence for intrinsic low temperature magnetism in SmB\textsubscript{6} \cite{gheidi_intrinsic_2019}.  Thus, it is reasonable to expect an additional $T^3 \ln{(T/T^{\star})}$ contribution due to spin fluctuations, as previously observed in other Kondo systems~\cite{ikeda_quenching_1991}, heavy fermions \cite{stewart_possibility_1984} and other electron mass-enhanced metals  \cite{brinkman_spin-fluctuation_1968}.    In SmB\textsubscript{6}, the $T^3\ln(T)$ term has been used to model the dependence of the low temperature specific heat of SmB\textsubscript{6} on carbon doping \cite{phelan_correlation_2014} and (La, Yb) rare earth substitution \cite{orendac_isosbestic_2017}.  Specific heat measurements in a field can therefore provide a critical test: if spin fluctuations are the source of the zero field $T^3 \ln{(T/T^{\star})}$ contribution  and mass enhancement $m^{\star}/m$, that enhancement should be  significantly reduced for fields greater than or on the order of \(H^{\ast}=\frac{k_BT^{\ast}}{\mu_B}\) where $k_BT^*$ is a characteristic energy for spin fluctuations. This reduction results in a decrease in the quasi-particle enhanced effective mass ratio $m^{\star}/m$ and thus $\gamma(H)$,  
which should be proportional to \((H/H^\star)^2\) at low fields \cite{hertel_effect_1980, beal-monod_field_1983}. For $ T^*$ = 15 K, we expect $H^* = 23 \text{T}$.

Our observations at both low and high temperatures are consistent with the previously observed behavior discussed above.  Consistent with this expectation, we find that  $C/T$ for all measured temperatures ($T\leq 1K$) begins to significantly decrease above 18 T, leveling off above 22 T; the initial field dependence is proportional to $(H/H^*)^2$ (Fig.~\ref{fig:zerofield} inset). A low field Schottky peak around 1-2 T arises from the magnetic impurities present in the sample, as expected; we attribute a second peak seen in higher fields around 12 - 15 T to the experimentally observed suppression of the gap between the in-gap states and the conduction band \cite{caldwell_high-field_2007} by a magnetic field on the order of 14 T (see supplemental for details).  


We turn now to the oscillatory component of the specific heat. Magnetoquantum oscillations in thermodynamic quantities such as magnetization and specific heat arise from oscillations in the thermodynamic potential, the free energy minus the chemical potential of the system, \(\tilde{\Omega} = f_T(z) \tilde{\Omega}_0\) where in the L-K model \cite{shoenberg_magnetic_1984}, \(f_T\left(z\right)= z / \sinh{z} \) represents the reduction in oscillation amplitude at finite temperature due to thermal smearing of the Fermi surface and \( \tilde{\Omega}_0\) is the zero temperature thermodynamic potential.  Here 
\begin{equation}
\label{eq:z}
    z \equiv   \pi^2 p \left(\frac{m^{\star}}{m} \right)\left(\frac{m}{m_e} \right) \left(\frac{k_B}{ \mu_B}\right)\frac{T}{H} ,
\end{equation}
where \(m^{\star}\)is the quasi-particle interaction enhanced mass, \(m\) is the band mass, \(m_e\) is the bare electron mass, and $p$ is an integer denoting the harmonic. 

Unlike typical specific heat measurements, the contribution due to cyclotron motion in a magnetic or gauge field will be subject to a damping term related to quasiparticle scattering.  Introducing quasiparticle scattering results in an additional damping term in the thermodynamic potential so that  \(\tilde{\Omega} =  f_T(z) f_D\ \tilde{\Omega}_0\) where  $f_D = e^{-p H_D / H} $ and 
\begin{equation}
\label{eq:z_D}
H_D  = \pi^2  \left(\frac{k_B }{\mu_B}\right)T_D^{\star}
\end{equation}
where $T_D^{\star}$ is the effective mass Dingle temperature, related to the original band mass Dingle temperature $T_D$  by  $T_D^{\star} = \frac{m^{\star}}{m_e } T_D$ \cite{dingle_magnetic_1952-1, dingle_magnetic_1952-2, shoenberg_magnetic_1984}.

Normally such MQOs arise from the motion of charge carriers and here we are asking, irrespective of the nature of coupling to a gauge field, is there evidence for MQOs in the specific heat and thus the density of states?  The oscillatory component of the specific heat depends on the second derivative of the thermodynamic potential,
\begin{equation}
\label{eq:c_osc}
\tilde{C} = -T \frac{\partial}{\partial T}\left(\frac{\partial \tilde{\Omega}}{\partial T}\right) = -\frac{1}{T} z^2 f_T^{\prime \prime}(z) f_D\ \tilde{\Omega}_0
\end{equation}
which, unlike the oscillatory component of the magnetization, goes to zero at a nonzero finite field corresponding to \(z\approx 1.61\).  We can use the appearance of this node not only to validate the measurement but also to determine the effective mass enhancement directly from the L-K model, without need for a study of $T$ dependence \cite{bondarenko_first_2001}. 

Magneto-quantum oscillations in $C(T)$ are  predicted to be, intrinsically, a much smaller fraction of the total signal than the electronic component of $C(T)$ in zero field \cite{shoenberg_magnetic_1984}. This small signal creates a novel challenge for low signal-to-noise measurements such as ours. To avoid introducing false peaks in the Fourier spectrum, as well as systematic errors in oscillation peak frequency identifications, we use a more general non-uniform discrete Fourier transform (NDFT), the Lomb-Scargel (LS) method \cite{vanderplas_understanding_2018}. This method, widely used in astronomy, avoids the aliasing errors that would otherwise be introduced by the usual interpolation techniques for generating the uniform-in-$1/H$ data sets needed for accurate MQO frequency determination using Fourier transforms. An important starting point for implementing the LS method is the subtraction of a uniform $C(H)$ background and, to avoid introducing low-frequency terms.  We therefore avoided using a multi term polynomial, and instead subtracted a sigmoid function least squares fit before carrying out the frequency analysis.  A discussion of our use of the Lomb-Scargle method and the applied peak selection criteria is presented in the supplemental material.

\begin{figure}[htb!]
\begin{center}
\includegraphics[width=\columnwidth]{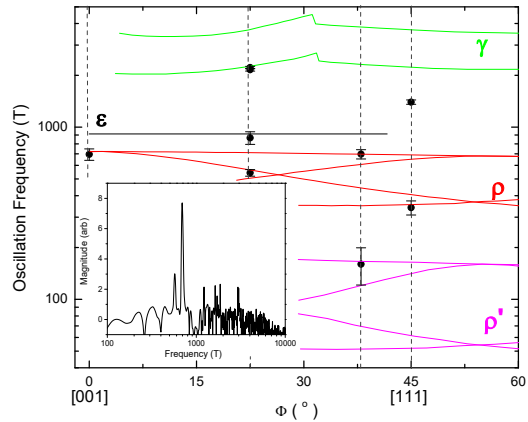}
\caption{{\label{fig:SmB6_angle_dependence} Angular dependence of the oscillation frequency for a 1.511 mg flux grown sample of SmB\textsubscript{6} (solid points). Solid lines correspond to predicted angle-dependent oscillation frequencies for SmB\textsubscript{6} \cite{hartstein_fermi_2018}. Band labels $\rho$, $\rho^{\prime}$, $\epsilon$, and $\gamma$ are as presented in \cite{hartstein_fermi_2018}.  Inset: Sample weighted periodogram for $\phi = 0^o$. The main peak corresponds to an oscillation frequency of 695 T.
}
}
\end{center}
\end{figure}
As a test of our method, we first measured the oscillatory specific heat of a 0.085 $\mu$g sample of LaB\textsubscript{6} at temperatures ranging from 0.1 K-1.0 K in fields up to 12 T.
Applying the Lomb-Scargle frequency analysis described above, we resolved frequency peaks at F = 847($\pm 8$), 1697($\pm 18$), 3228($\pm 15$), 7866($\pm 16$), and 15732($\pm 21$) T, in good agreement with
previously reported values of 845, 1690, 3220, 7800, and 15600 T based on dHvA measurements \cite{ishizawa_haas-van_1977} (as shown in the supplemental material). The effective masses were also the same as determined by dHvA, ranging from 0.066 to 0.65 $m_e$ for the respective bands.

For SmB\textsubscript{6}, we applied the same analysis method as used for LaB\textsubscript{6}.  We find magnetic field orientation-dependent frequencies in general agreement with previous results \cite{hartstein_fermi_2018}. In  Fig. \ref{fig:SmB6_angle_dependence} we show a comparison of measured frequencies  determined from the data after background subtraction, $C_{res}(H)$ collected at 0.58 K  between 18 and 31 T with DFT predictions (green, black, red and pink lines \cite{tan_unconventional_2015}).  We are unable to clearly resolve the lowest expected oscillation frequencies but find approximately a 90\% fidelity agreement of the oscillation frequencies we do observe with those reported from de Haas-van Alphen (dHvA) measurements on float zone growth samples \cite{hartstein_fermi_2018}.

\begin{figure}[ht!]
\begin{center}
\includegraphics[width= 1.1\columnwidth]{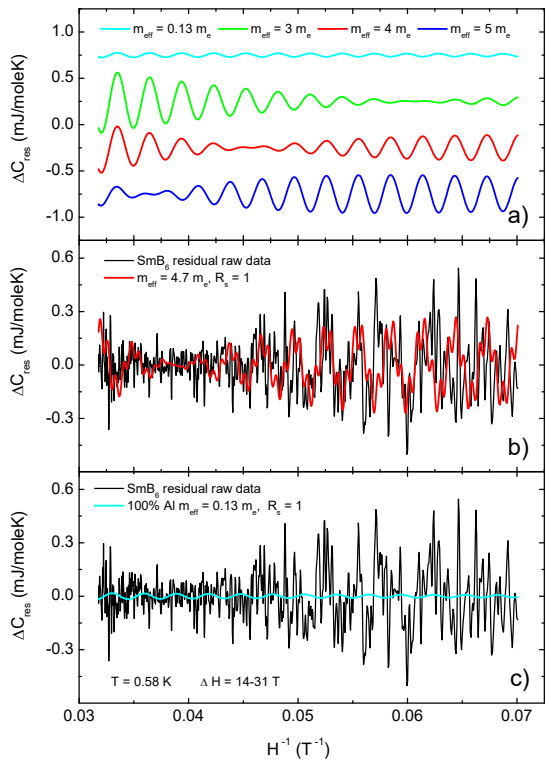}
\caption{{{\label{fig:LKfits}}(a) Predicted oscillatory specific heat for effective mass enhancement ratios of 0.13, 3.0, 4.0, and 5.0 respectively, assuming a sample temperature T = 0.58K and an impurity scattering Dingle temperature $T_D = 1$ K. Traces have been offset vertically for clarity. (b) Residual specific heat vs $H^{-1}$ at T = 0.58K and $\Phi =45^o$. The red curve is our best fit of the L-K model to the data, assuming an effective mass of $4.7 m_e$, $H_D = 10.9$ T, and no attenuation due to spin-scattering ($R_S = 1$.) A $\pi$ phase shift was introduced at 26 T to best fit the data, a factor needed for high field QO measurements especially in the presence of magnetic breakdown \cite{reifenberger_1979} (c) Comparison of the data with the predictions of the L-K model for a 100\% Al sample (effective mass of $0.13m_e$, $T_D = 1$ K , which gives $H_D = 19.5$ T, and $R_S = 1$).}}
\end{center}
\end{figure}
One cause for caution is that flux grown samples often possess Al inclusions, and  torque measurements have shown inclusions can produce MQOs at frequencies similar to those expected for SmB\textsubscript{6} \cite{thomas_quantum_2019}.  In Fig.~\ref{fig:LKfits}(c) we therefore show a comparison of the the corresponding oscillation amplitude and magnetic field dependence expected for an aluminum sample, using the known oscillation frequencies and effective masses of Al \cite{larson_low-field_1967}.  We see here  that even at the 100\% level (pure Al), we are unable to account for the amplitude of the MQOs we see in the specific heat.  In our flux grown samples, the absence of a discernible jump in the zero-field electronic specific heat of 1\% or greater at the Al superconducting transition temperature of 1.163 K places an upper limit on the actual Al percentage of less than 5\%.  We believe that this is a critical test since, if Al inclusions are producing MQOs, then they must arise from high quality crystalline material.

Further confidence in our analysis comes from our observation of a  node in the magnetic-field-dependence of the  MQOs \cite{shoenberg_magnetic_1984, bondarenko_first_2001} at 0.58 K.  First, in Fig. 3(a), we show  the expected magnetic field dependence of the oscillatory specific heat profile on effective mass for an oscillation frequency of 341 tesla at 0.58 K. Second, in Fig. 3(b), we show the corresponding best fits of the L-K model to a representative data trace at 0.58 K for the 1.511 mg sample of SmB\textsubscript{6} between 14 - 31 T measured using the larger microcalorimeter \cite{fortune_top-loading_2014}.  The MQOs observed in the specific heat have a node at approximately 26 T, indicating an average enhanced effective mass ratio of 4.7. As mentioned above, we expect this value to remain constant for fields $\geq 23 $ T.  Finally, in Fig. 3(c), we compare the same data trace with that expected for a 100\% aluminum sample of the same mass. The location of this node at 26 T is incompatible with that which would be observed for aluminum, given the known effective masses for Al. A discussion of the fitting procedure is outlined in the supplemental material.

\begin{figure}[htb!]
\includegraphics[width=1.1\columnwidth]{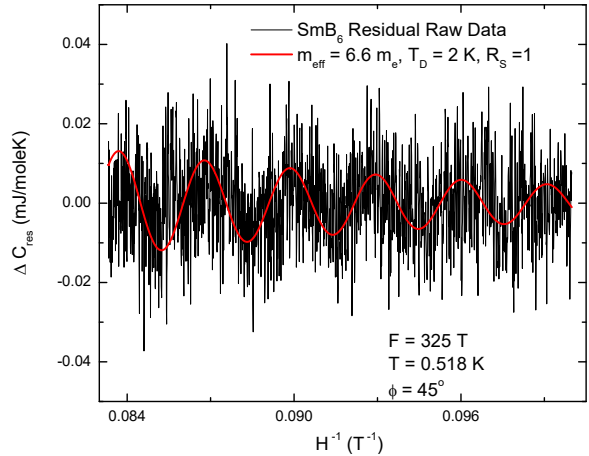}
\caption{\label{fig:membranenanocalorimeter}  Residual specific heat vs $H^{-1}$ at T = 0.518 K for $\Phi = 45^o$. The black curve is the raw data post background subtraction.~ The red curve is a sample L-K fit using the sample parameters as m\textsubscript{eff} = 6.6 m\textsubscript{e} T\textsubscript{D} = 2K, and R\textsubscript{S} = 1}
\end{figure}

For confirmation of the observed heavy MQOs in $C(H)$ of SmB\textsubscript{6}, a second set of measurements were made on a 0.126 $\mu$g Al-flux grown sample using a high-resolution membrane nanocalorimeter \cite{tagliati_differential_2012}, as shown in Fig.~\ref{fig:membranenanocalorimeter} for a field sweep at 0.52 K for $H \parallel [111]$. We find observable MQOs between 10-12 T with a  mass enhancement of 6.6 $m_e$, slightly enhanced from the 4.7 $m_e$ value observed at similar temperature at high field.

We note that while the MQO frequencies in $C$ are in good agreement with those obtained from magnetic torque measurements, the mass enhancements are not.  
We find effective mass enhancements \(\frac{m^{\ast}}{m}\) ranging from 4.5 to 6.6  from fits of the L-K model to the data for both data sets, in contrast with values ranging from \(\frac{m^{\ast}}{m}\) = 0.1 - 1.0   found from torque magnetometry  \cite{li_two-dimensional_2014, tan_unconventional_2015, xiang_bulk_2017, hartstein_fermi_2018}.  Nevertheless, our larger mass enhancements are consistent with those determined from the node in oscillation amplitude observed at 0.58 K and the large mass enhancements found in our fit to the zero-field data.  Finally, the larger effective masses are consistent with a recent first-principles, parameter-free all-electron electronic-structure model for SmB\textsubscript{6}  (\(\frac{m^{\ast}}{m}\) = 2.0 - 22.0 depending on the band) \cite{zhang_understanding_2020}. 

One possible theoretical explanation for the discrepancy in effective mass values observed by specific heat and magnetic torque would be the simultaneous existence of light and heavy quasiparticle masses, as has been proposed for SmB$_6$   \cite{harrison_highly_2018a}.  In this theoretical model, the MQOs arise when a highly asymmetric nodal semimetal forms at low temperature with carriers populated from disorder-induced in-gap states in small-gap Kondo insulators \cite{shen_quantum_2018}. Whether this theory allows the formation of charge neutral excitations is not clear to us but in any case, it would be interesting if the theory were to be extended to include a  calculation of the oscillatory specific heat,  so as to enable a more direct comparison with our results.  Additionally, recent experimental studies on the Kondo insulator YbB\textsubscript{12} suggest a two-fluid picture for the origin of the observed MQO profile in which neutral quasiparticles coexist with charged fermions \cite{Xiang_YbB12}. Finally it has also been shown in recent theoretical work that neutral quasiparticles arise naturally in mixed valence systems as Majorana excitations \cite{varma_majorana_2020}.  Such excitations would exhibit no charge transport in linear response, but would indeed show MQOs in magnetization, as well as specific heat, consistent with our observations.  Further studies to accurately determine spin-splitting attenuation factors are required to support or oppose these claims.

In conclusion we have resolved MQOs in the high field residual specific heat of SmB\textsubscript{6} that show good agreement with theoretical expectations for the dependence of oscillation frequency on crystallographic orientation for SmB\textsubscript{6}, even though the parameters needed to describe the observed specific heat oscillations within L-K theory  indicate much larger masses than those previously determined from dHvA measurements.  In future measurements, we hope to use still higher sensitivity calorimeters to measure $C(H,T)$ vs $\phi$ systematically in high fields to probe for light and heavy effective masses in high quality float- and flux-grown samples at low temperatures, with a goal of the direct observation of oscillations in the specific heat at high magnetic fields.

\begin{acknowledgements}

We would like to acknowledge the contributions of Dale Renfrow of the Smith College Center for Design and Fabrication and 
Ju-Hyun Park, William Coniglio, and Ali Bangura  of NHMFL for technical support.
The samples were grown at Los Alamos National Laboratory by P. F. Rosa.  
This work was also supported in part by the U.S. Department of Energy Office of Basic Energy Science, Division of Condensed Matter Physics grant DE-SC0017862 (P.G.L. and A.P.R.).  A portion of this work was performed at the National High Magnetic Field Laboratory, which is supported by the National Science Foundation Cooperative Agreement No. DMR-1644779 and the State of Florida.
\end{acknowledgements}

\bibliography{SmB6main}

\providecommand{\noopsort}[1]{}\providecommand{\singleletter}[1]{#1}%
\begin{thebibliography}{46}%
\makeatletter
\providecommand \@ifxundefined [1]{%
 \@ifx{#1\undefined}
}%
\providecommand \@ifnum [1]{%
 \ifnum #1\expandafter \@firstoftwo
 \else \expandafter \@secondoftwo
 \fi
}%
\providecommand \@ifx [1]{%
 \ifx #1\expandafter \@firstoftwo
 \else \expandafter \@secondoftwo
 \fi
}%
\providecommand \natexlab [1]{#1}%
\providecommand \enquote  [1]{``#1''}%
\providecommand \bibnamefont  [1]{#1}%
\providecommand \bibfnamefont [1]{#1}%
\providecommand \citenamefont [1]{#1}%
\providecommand \href@noop [0]{\@secondoftwo}%
\providecommand \href [0]{\begingroup \@sanitize@url \@href}%
\providecommand \@href[1]{\@@startlink{#1}\@@href}%
\providecommand \@@href[1]{\endgroup#1\@@endlink}%
\providecommand \@sanitize@url [0]{\catcode `\\12\catcode `\$12\catcode
  `\&12\catcode `\#12\catcode `\^12\catcode `\_12\catcode `\%12\relax}%
\providecommand \@@startlink[1]{}%
\providecommand \@@endlink[0]{}%
\providecommand \url  [0]{\begingroup\@sanitize@url \@url }%
\providecommand \@url [1]{\endgroup\@href {#1}{\urlprefix }}%
\providecommand \urlprefix  [0]{URL }%
\providecommand \Eprint [0]{\href }%
\providecommand \doibase [0]{https://doi.org/}%
\providecommand \selectlanguage [0]{\@gobble}%
\providecommand \bibinfo  [0]{\@secondoftwo}%
\providecommand \bibfield  [0]{\@secondoftwo}%
\providecommand \translation [1]{[#1]}%
\providecommand \BibitemOpen [0]{}%
\providecommand \bibitemStop [0]{}%
\providecommand \bibitemNoStop [0]{.\EOS\space}%
\providecommand \EOS [0]{\spacefactor3000\relax}%
\providecommand \BibitemShut  [1]{\csname bibitem#1\endcsname}%
\let\auto@bib@innerbib\@empty
\bibitem [{\citenamefont {Hartstein}\ \emph {et~al.}(2020)\citenamefont
  {Hartstein}, \citenamefont {Liu}, \citenamefont {Hsu}, \citenamefont {Tan},
  \citenamefont {Ciomaga~Hatnean}, \citenamefont {Balakrishnan},\ and\
  \citenamefont {Sebastian}}]{hartstein_intrinsic_2020}%
  \BibitemOpen
  \bibfield  {author} {\bibinfo {author} {\bibfnamefont {M.}~\bibnamefont
  {Hartstein}}, \bibinfo {author} {\bibfnamefont {H.}~\bibnamefont {Liu}},
  \bibinfo {author} {\bibfnamefont {Y.-T.}\ \bibnamefont {Hsu}}, \bibinfo
  {author} {\bibfnamefont {B.~S.}\ \bibnamefont {Tan}}, \bibinfo {author}
  {\bibfnamefont {M.}~\bibnamefont {Ciomaga~Hatnean}}, \bibinfo {author}
  {\bibfnamefont {G.}~\bibnamefont {Balakrishnan}},\ and\ \bibinfo {author}
  {\bibfnamefont {S.~E.}\ \bibnamefont {Sebastian}},\ }\href@noop {} {\bibfield
   {journal} {\bibinfo  {journal} {iScience}\ }\textbf {\bibinfo {volume}
  {23}},\ \bibinfo {pages} {101632} (\bibinfo {year} {2020})}\BibitemShut
  {NoStop}%
\bibitem [{\citenamefont {Wolgast}\ \emph {et~al.}(2017)\citenamefont
  {Wolgast}, \citenamefont {Eo}, \citenamefont {Sun}, \citenamefont {Kurdak},
  \citenamefont {Balakirev}, \citenamefont {Jaime}, \citenamefont {Kim},\ and\
  \citenamefont {Fisk}}]{Wolgast_2017}%
  \BibitemOpen
  \bibfield  {author} {\bibinfo {author} {\bibfnamefont {S.}~\bibnamefont
  {Wolgast}}, \bibinfo {author} {\bibfnamefont {Y.~S.}\ \bibnamefont {Eo}},
  \bibinfo {author} {\bibfnamefont {K.}~\bibnamefont {Sun}}, \bibinfo {author}
  {\bibfnamefont {i.~m.~c.}\ \bibnamefont {Kurdak}}, \bibinfo {author}
  {\bibfnamefont {F.~F.}\ \bibnamefont {Balakirev}}, \bibinfo {author}
  {\bibfnamefont {M.}~\bibnamefont {Jaime}}, \bibinfo {author} {\bibfnamefont
  {D.-J.}\ \bibnamefont {Kim}},\ and\ \bibinfo {author} {\bibfnamefont
  {Z.}~\bibnamefont {Fisk}},\ }\href
  {https://doi.org/10.1103/PhysRevB.95.245112} {\bibfield  {journal} {\bibinfo
  {journal} {Phys. Rev. B}\ }\textbf {\bibinfo {volume} {95}},\ \bibinfo
  {pages} {245112} (\bibinfo {year} {2017})}\BibitemShut {NoStop}%
\bibitem [{\citenamefont {Li}\ \emph {et~al.}(2014)\citenamefont {Li},
  \citenamefont {Xiang}, \citenamefont {Yu}, \citenamefont {Asaba},
  \citenamefont {Lawson}, \citenamefont {Cai}, \citenamefont {Tinsman},
  \citenamefont {Berkley}, \citenamefont {Wolgast}, \citenamefont {Eo},
  \citenamefont {Kim}, \citenamefont {Kurdak}, \citenamefont {Allen},
  \citenamefont {Sun}, \citenamefont {Chen}, \citenamefont {Wang},
  \citenamefont {Fisk},\ and\ \citenamefont {Li}}]{li_two-dimensional_2014}%
  \BibitemOpen
  \bibfield  {author} {\bibinfo {author} {\bibfnamefont {G.}~\bibnamefont
  {Li}}, \bibinfo {author} {\bibfnamefont {Z.}~\bibnamefont {Xiang}}, \bibinfo
  {author} {\bibfnamefont {F.}~\bibnamefont {Yu}}, \bibinfo {author}
  {\bibfnamefont {T.}~\bibnamefont {Asaba}}, \bibinfo {author} {\bibfnamefont
  {B.}~\bibnamefont {Lawson}}, \bibinfo {author} {\bibfnamefont
  {P.}~\bibnamefont {Cai}}, \bibinfo {author} {\bibfnamefont {C.}~\bibnamefont
  {Tinsman}}, \bibinfo {author} {\bibfnamefont {A.}~\bibnamefont {Berkley}},
  \bibinfo {author} {\bibfnamefont {S.}~\bibnamefont {Wolgast}}, \bibinfo
  {author} {\bibfnamefont {Y.~S.}\ \bibnamefont {Eo}}, \bibinfo {author}
  {\bibfnamefont {D.-J.}\ \bibnamefont {Kim}}, \bibinfo {author} {\bibfnamefont
  {C.}~\bibnamefont {Kurdak}}, \bibinfo {author} {\bibfnamefont {J.~W.}\
  \bibnamefont {Allen}}, \bibinfo {author} {\bibfnamefont {K.}~\bibnamefont
  {Sun}}, \bibinfo {author} {\bibfnamefont {X.~H.}\ \bibnamefont {Chen}},
  \bibinfo {author} {\bibfnamefont {Y.~Y.}\ \bibnamefont {Wang}}, \bibinfo
  {author} {\bibfnamefont {Z.}~\bibnamefont {Fisk}},\ and\ \bibinfo {author}
  {\bibfnamefont {L.}~\bibnamefont {Li}},\ }\href@noop {} {\bibfield  {journal}
  {\bibinfo  {journal} {Science}\ }\textbf {\bibinfo {volume} {346}},\ \bibinfo
  {pages} {1208} (\bibinfo {year} {2014})}\BibitemShut {NoStop}%
\bibitem [{\citenamefont {Tan}\ \emph {et~al.}(2015)\citenamefont {Tan},
  \citenamefont {Hsu}, \citenamefont {Zeng}, \citenamefont {Hatnean},
  \citenamefont {Harrison}, \citenamefont {Zhu}, \citenamefont {Hartstein},
  \citenamefont {Kiourlappou}, \citenamefont {Srivastava}, \citenamefont
  {Johannes}, \citenamefont {Murphy}, \citenamefont {Park}, \citenamefont
  {Balicas}, \citenamefont {Lonzarich}, \citenamefont {Balakrishnan},\ and\
  \citenamefont {Sebastian}}]{tan_unconventional_2015}%
  \BibitemOpen
  \bibfield  {author} {\bibinfo {author} {\bibfnamefont {B.~S.}\ \bibnamefont
  {Tan}}, \bibinfo {author} {\bibfnamefont {Y.-T.}\ \bibnamefont {Hsu}},
  \bibinfo {author} {\bibfnamefont {B.}~\bibnamefont {Zeng}}, \bibinfo {author}
  {\bibfnamefont {M.~C.}\ \bibnamefont {Hatnean}}, \bibinfo {author}
  {\bibfnamefont {N.}~\bibnamefont {Harrison}}, \bibinfo {author}
  {\bibfnamefont {Z.}~\bibnamefont {Zhu}}, \bibinfo {author} {\bibfnamefont
  {M.}~\bibnamefont {Hartstein}}, \bibinfo {author} {\bibfnamefont
  {M.}~\bibnamefont {Kiourlappou}}, \bibinfo {author} {\bibfnamefont
  {A.}~\bibnamefont {Srivastava}}, \bibinfo {author} {\bibfnamefont {M.~D.}\
  \bibnamefont {Johannes}}, \bibinfo {author} {\bibfnamefont {T.~P.}\
  \bibnamefont {Murphy}}, \bibinfo {author} {\bibfnamefont {J.-H.}\
  \bibnamefont {Park}}, \bibinfo {author} {\bibfnamefont {L.}~\bibnamefont
  {Balicas}}, \bibinfo {author} {\bibfnamefont {G.~G.}\ \bibnamefont
  {Lonzarich}}, \bibinfo {author} {\bibfnamefont {G.}~\bibnamefont
  {Balakrishnan}},\ and\ \bibinfo {author} {\bibfnamefont {S.~E.}\ \bibnamefont
  {Sebastian}},\ }\href@noop {} {\bibfield  {journal} {\bibinfo  {journal}
  {Science}\ }\textbf {\bibinfo {volume} {349}},\ \bibinfo {pages} {287}
  (\bibinfo {year} {2015})}\BibitemShut {NoStop}%
\bibitem [{\citenamefont {Xiang}\ \emph {et~al.}(2017)\citenamefont {Xiang},
  \citenamefont {Lawson}, \citenamefont {Asaba}, \citenamefont {Tinsman},
  \citenamefont {Chen}, \citenamefont {Shang}, \citenamefont {Chen},\ and\
  \citenamefont {Li}}]{xiang_bulk_2017}%
  \BibitemOpen
  \bibfield  {author} {\bibinfo {author} {\bibfnamefont {Z.}~\bibnamefont
  {Xiang}}, \bibinfo {author} {\bibfnamefont {B.}~\bibnamefont {Lawson}},
  \bibinfo {author} {\bibfnamefont {T.}~\bibnamefont {Asaba}}, \bibinfo
  {author} {\bibfnamefont {C.}~\bibnamefont {Tinsman}}, \bibinfo {author}
  {\bibfnamefont {L.}~\bibnamefont {Chen}}, \bibinfo {author} {\bibfnamefont
  {C.}~\bibnamefont {Shang}}, \bibinfo {author} {\bibfnamefont
  {X.}~\bibnamefont {Chen}},\ and\ \bibinfo {author} {\bibfnamefont
  {L.}~\bibnamefont {Li}},\ }\href@noop {} {\bibfield  {journal} {\bibinfo
  {journal} {Physical Review X}\ }\textbf {\bibinfo {volume} {7}},\ \bibinfo
  {pages} {031054} (\bibinfo {year} {2017})}\BibitemShut {NoStop}%
\bibitem [{\citenamefont {Hartstein}\ \emph {et~al.}(2018)\citenamefont
  {Hartstein}, \citenamefont {Toews}, \citenamefont {Hsu}, \citenamefont
  {Zeng}, \citenamefont {Chen}, \citenamefont {Hatnean}, \citenamefont {Zhang},
  \citenamefont {Nakamura}, \citenamefont {Padgett}, \citenamefont
  {Rodway-Gant}, \citenamefont {Berk}, \citenamefont {Kingston}, \citenamefont
  {Zhang}, \citenamefont {Chan}, \citenamefont {Yamashita}, \citenamefont
  {Sakakibara}, \citenamefont {Takano}, \citenamefont {Park}, \citenamefont
  {Balicas}, \citenamefont {Harrison}, \citenamefont {Shitsevalova},
  \citenamefont {Balakrishnan}, \citenamefont {Lonzarich}, \citenamefont
  {Hill}, \citenamefont {Sutherland},\ and\ \citenamefont
  {Sebastian}}]{hartstein_fermi_2018}%
  \BibitemOpen
  \bibfield  {author} {\bibinfo {author} {\bibfnamefont {M.}~\bibnamefont
  {Hartstein}}, \bibinfo {author} {\bibfnamefont {W.~H.}\ \bibnamefont
  {Toews}}, \bibinfo {author} {\bibfnamefont {Y.-T.}\ \bibnamefont {Hsu}},
  \bibinfo {author} {\bibfnamefont {B.}~\bibnamefont {Zeng}}, \bibinfo {author}
  {\bibfnamefont {X.}~\bibnamefont {Chen}}, \bibinfo {author} {\bibfnamefont
  {M.~C.}\ \bibnamefont {Hatnean}}, \bibinfo {author} {\bibfnamefont {Q.~R.}\
  \bibnamefont {Zhang}}, \bibinfo {author} {\bibfnamefont {S.}~\bibnamefont
  {Nakamura}}, \bibinfo {author} {\bibfnamefont {A.~S.}\ \bibnamefont
  {Padgett}}, \bibinfo {author} {\bibfnamefont {G.}~\bibnamefont
  {Rodway-Gant}}, \bibinfo {author} {\bibfnamefont {J.}~\bibnamefont {Berk}},
  \bibinfo {author} {\bibfnamefont {M.~K.}\ \bibnamefont {Kingston}}, \bibinfo
  {author} {\bibfnamefont {G.~H.}\ \bibnamefont {Zhang}}, \bibinfo {author}
  {\bibfnamefont {M.~K.}\ \bibnamefont {Chan}}, \bibinfo {author}
  {\bibfnamefont {S.}~\bibnamefont {Yamashita}}, \bibinfo {author}
  {\bibfnamefont {T.}~\bibnamefont {Sakakibara}}, \bibinfo {author}
  {\bibfnamefont {Y.}~\bibnamefont {Takano}}, \bibinfo {author} {\bibfnamefont
  {J.-H.}\ \bibnamefont {Park}}, \bibinfo {author} {\bibfnamefont
  {L.}~\bibnamefont {Balicas}}, \bibinfo {author} {\bibfnamefont
  {N.}~\bibnamefont {Harrison}}, \bibinfo {author} {\bibfnamefont
  {N.}~\bibnamefont {Shitsevalova}}, \bibinfo {author} {\bibfnamefont
  {G.}~\bibnamefont {Balakrishnan}}, \bibinfo {author} {\bibfnamefont {G.~G.}\
  \bibnamefont {Lonzarich}}, \bibinfo {author} {\bibfnamefont {R.~W.}\
  \bibnamefont {Hill}}, \bibinfo {author} {\bibfnamefont {M.}~\bibnamefont
  {Sutherland}},\ and\ \bibinfo {author} {\bibfnamefont {S.~E.}\ \bibnamefont
  {Sebastian}},\ }\href@noop {} {\bibfield  {journal} {\bibinfo  {journal}
  {Nature Physics}\ }\textbf {\bibinfo {volume} {14}},\ \bibinfo {pages} {166}
  (\bibinfo {year} {2018})}\BibitemShut {NoStop}%
\bibitem [{\citenamefont {Thomas}\ \emph {et~al.}(2019)\citenamefont {Thomas},
  \citenamefont {Ding}, \citenamefont {Ronning}, \citenamefont {Zapf},
  \citenamefont {Thompson}, \citenamefont {Fisk}, \citenamefont {Xia},\ and\
  \citenamefont {Rosa}}]{thomas_quantum_2019}%
  \BibitemOpen
  \bibfield  {author} {\bibinfo {author} {\bibfnamefont {S.}~\bibnamefont
  {Thomas}}, \bibinfo {author} {\bibfnamefont {X.}~\bibnamefont {Ding}},
  \bibinfo {author} {\bibfnamefont {F.}~\bibnamefont {Ronning}}, \bibinfo
  {author} {\bibfnamefont {V.}~\bibnamefont {Zapf}}, \bibinfo {author}
  {\bibfnamefont {J.}~\bibnamefont {Thompson}}, \bibinfo {author}
  {\bibfnamefont {Z.}~\bibnamefont {Fisk}}, \bibinfo {author} {\bibfnamefont
  {J.}~\bibnamefont {Xia}},\ and\ \bibinfo {author} {\bibfnamefont
  {P.}~\bibnamefont {Rosa}},\ }\href@noop {} {\bibfield  {journal} {\bibinfo
  {journal} {Physical Review Letters}\ }\textbf {\bibinfo {volume} {122}},\
  \bibinfo {pages} {166401} (\bibinfo {year} {2019})}\BibitemShut {NoStop}%
\bibitem [{\citenamefont {Li}\ \emph {et~al.}(2020)\citenamefont {Li},
  \citenamefont {Sun}, \citenamefont {Kurdak},\ and\ \citenamefont
  {Allen}}]{li_emergent_2020}%
  \BibitemOpen
  \bibfield  {author} {\bibinfo {author} {\bibfnamefont {L.}~\bibnamefont
  {Li}}, \bibinfo {author} {\bibfnamefont {K.}~\bibnamefont {Sun}}, \bibinfo
  {author} {\bibfnamefont {C.}~\bibnamefont {Kurdak}},\ and\ \bibinfo {author}
  {\bibfnamefont {J.~W.}\ \bibnamefont {Allen}},\ }\href
  {https://doi.org/10.1038/s42254-020-0210-8} {\bibfield  {journal} {\bibinfo
  {journal} {Nature Reviews Physics}\ }\textbf {\bibinfo {volume} {2}},\
  \bibinfo {pages} {463} (\bibinfo {year} {2020})}\BibitemShut {NoStop}%
\bibitem [{\citenamefont {Knolle}\ and\ \citenamefont
  {Cooper}(2015)}]{Knolle_2015}%
  \BibitemOpen
  \bibfield  {author} {\bibinfo {author} {\bibfnamefont {J.}~\bibnamefont
  {Knolle}}\ and\ \bibinfo {author} {\bibfnamefont {N.~R.}\ \bibnamefont
  {Cooper}},\ }\href {https://doi.org/10.1103/PhysRevLett.115.146401}
  {\bibfield  {journal} {\bibinfo  {journal} {Phys. Rev. Lett.}\ }\textbf
  {\bibinfo {volume} {115}},\ \bibinfo {pages} {146401} (\bibinfo {year}
  {2015})}\BibitemShut {NoStop}%
\bibitem [{\citenamefont {Knolle}\ and\ \citenamefont
  {Cooper}(2017)}]{Knolle_2017}%
  \BibitemOpen
  \bibfield  {author} {\bibinfo {author} {\bibfnamefont {J.}~\bibnamefont
  {Knolle}}\ and\ \bibinfo {author} {\bibfnamefont {N.~R.}\ \bibnamefont
  {Cooper}},\ }\href {https://doi.org/10.1103/PhysRevLett.118.096604}
  {\bibfield  {journal} {\bibinfo  {journal} {Phys. Rev. Lett.}\ }\textbf
  {\bibinfo {volume} {118}},\ \bibinfo {pages} {096604} (\bibinfo {year}
  {2017})}\BibitemShut {NoStop}%
\bibitem [{\citenamefont {Chowdhury}\ \emph {et~al.}(2018)\citenamefont
  {Chowdhury}, \citenamefont {Sodemann},\ and\ \citenamefont
  {Senthil}}]{Chowdhury_2018}%
  \BibitemOpen
  \bibfield  {author} {\bibinfo {author} {\bibfnamefont {D.}~\bibnamefont
  {Chowdhury}}, \bibinfo {author} {\bibfnamefont {I.}~\bibnamefont
  {Sodemann}},\ and\ \bibinfo {author} {\bibfnamefont {T.}~\bibnamefont
  {Senthil}},\ }\href {https://doi.org/10.1038/s41467-018-04163-2} {\bibfield
  {journal} {\bibinfo  {journal} {Nature Communications}\ }\textbf {\bibinfo
  {volume} {9}},\ \bibinfo {pages} {1766} (\bibinfo {year} {2018})}\BibitemShut
  {NoStop}%
\bibitem [{\citenamefont {Sullivan}\ and\ \citenamefont
  {Seidel}(1968)}]{sullivan_steady-state_1968}%
  \BibitemOpen
  \bibfield  {author} {\bibinfo {author} {\bibfnamefont {P.~F.}\ \bibnamefont
  {Sullivan}}\ and\ \bibinfo {author} {\bibfnamefont {G.}~\bibnamefont
  {Seidel}},\ }\href@noop {} {\bibfield  {journal} {\bibinfo  {journal}
  {Physical Review}\ }\textbf {\bibinfo {volume} {173}},\ \bibinfo {pages}
  {679} (\bibinfo {year} {1968})}\BibitemShut {NoStop}%
\bibitem [{\citenamefont {Shoenberg}(1984)}]{shoenberg_magnetic_1984}%
  \BibitemOpen
  \bibfield  {author} {\bibinfo {author} {\bibfnamefont {D.}~\bibnamefont
  {Shoenberg}},\ }\href@noop {} {\emph {\bibinfo {title} {Magnetic Oscillations
  in Metals}}}\ (\bibinfo  {publisher} {Cambridge University Press},\ \bibinfo
  {address} {Cambridge},\ \bibinfo {year} {1984})\BibitemShut {NoStop}%
\bibitem [{\citenamefont {Varma}(2020)}]{varma_majorana_2020}%
  \BibitemOpen
  \bibfield  {author} {\bibinfo {author} {\bibfnamefont {C.~M.}\ \bibnamefont
  {Varma}},\ }\href {https://doi.org/10.1103/PhysRevB.102.155145} {\bibfield
  {journal} {\bibinfo  {journal} {Phys. Rev. B}\ }\textbf {\bibinfo {volume}
  {102}},\ \bibinfo {pages} {155145} (\bibinfo {year} {2020})}\BibitemShut
  {NoStop}%
\bibitem [{\citenamefont {Fortune}\ \emph {et~al.}(1990)\citenamefont
  {Fortune}, \citenamefont {Brooks}, \citenamefont {Graf}, \citenamefont
  {Montambaux}, \citenamefont {Chiang}, \citenamefont {Perenboom},\ and\
  \citenamefont {Althof}}]{fortune_specific-heat_1990}%
  \BibitemOpen
  \bibfield  {author} {\bibinfo {author} {\bibfnamefont {N.~A.}\ \bibnamefont
  {Fortune}}, \bibinfo {author} {\bibfnamefont {J.~S.}\ \bibnamefont {Brooks}},
  \bibinfo {author} {\bibfnamefont {M.~J.}\ \bibnamefont {Graf}}, \bibinfo
  {author} {\bibfnamefont {G.}~\bibnamefont {Montambaux}}, \bibinfo {author}
  {\bibfnamefont {L.~Y.}\ \bibnamefont {Chiang}}, \bibinfo {author}
  {\bibfnamefont {J.~A. A.~J.}\ \bibnamefont {Perenboom}},\ and\ \bibinfo
  {author} {\bibfnamefont {D.}~\bibnamefont {Althof}},\ }\href
  {https://doi.org/10.1103/PhysRevLett.64.2054} {\bibfield  {journal} {\bibinfo
   {journal} {Physical Review Letters}\ }\textbf {\bibinfo {volume} {64}},\
  \bibinfo {pages} {2054} (\bibinfo {year} {1990})}\BibitemShut {NoStop}%
\bibitem [{\citenamefont {Bondarenko}\ \emph {et~al.}(2001)\citenamefont
  {Bondarenko}, \citenamefont {Uji}, \citenamefont {Terashima}, \citenamefont
  {Terakura}, \citenamefont {Tanaka}, \citenamefont {Maki}, \citenamefont
  {Yamada},\ and\ \citenamefont {Nakatsuji}}]{bondarenko_first_2001}%
  \BibitemOpen
  \bibfield  {author} {\bibinfo {author} {\bibfnamefont {V.~A.}\ \bibnamefont
  {Bondarenko}}, \bibinfo {author} {\bibfnamefont {S.}~\bibnamefont {Uji}},
  \bibinfo {author} {\bibfnamefont {T.}~\bibnamefont {Terashima}}, \bibinfo
  {author} {\bibfnamefont {C.}~\bibnamefont {Terakura}}, \bibinfo {author}
  {\bibfnamefont {S.}~\bibnamefont {Tanaka}}, \bibinfo {author} {\bibfnamefont
  {S.}~\bibnamefont {Maki}}, \bibinfo {author} {\bibfnamefont {J.}~\bibnamefont
  {Yamada}},\ and\ \bibinfo {author} {\bibfnamefont {S.}~\bibnamefont
  {Nakatsuji}},\ }\href {https://doi.org/10.1016/S0379-6779(00)01149-8}
  {\bibfield  {journal} {\bibinfo  {journal} {Synthetic Metals}\ }\textbf
  {\bibinfo {volume} {120}},\ \bibinfo {pages} {1039} (\bibinfo {year}
  {2001})}\BibitemShut {NoStop}%
\bibitem [{\citenamefont {Stankiewicz}\ \emph {et~al.}(2019)\citenamefont
  {Stankiewicz}, \citenamefont {Evangelisti}, \citenamefont {Rosa},
  \citenamefont {Schlottmann},\ and\ \citenamefont
  {Fisk}}]{PhysRevB.99.045138}%
  \BibitemOpen
  \bibfield  {author} {\bibinfo {author} {\bibfnamefont {J.}~\bibnamefont
  {Stankiewicz}}, \bibinfo {author} {\bibfnamefont {M.}~\bibnamefont
  {Evangelisti}}, \bibinfo {author} {\bibfnamefont {P.~F.~S.}\ \bibnamefont
  {Rosa}}, \bibinfo {author} {\bibfnamefont {P.}~\bibnamefont {Schlottmann}},\
  and\ \bibinfo {author} {\bibfnamefont {Z.}~\bibnamefont {Fisk}},\ }\href
  {https://doi.org/10.1103/PhysRevB.99.045138} {\bibfield  {journal} {\bibinfo
  {journal} {Phys. Rev. B}\ }\textbf {\bibinfo {volume} {99}},\ \bibinfo
  {pages} {045138} (\bibinfo {year} {2019})}\BibitemShut {NoStop}%
\bibitem [{\citenamefont {Tagliati}\ \emph {et~al.}(2012)\citenamefont
  {Tagliati}, \citenamefont {Krasnov},\ and\ \citenamefont
  {Rydh}}]{tagliati_differential_2012}%
  \BibitemOpen
  \bibfield  {author} {\bibinfo {author} {\bibfnamefont {S.}~\bibnamefont
  {Tagliati}}, \bibinfo {author} {\bibfnamefont {V.~M.}\ \bibnamefont
  {Krasnov}},\ and\ \bibinfo {author} {\bibfnamefont {A.}~\bibnamefont
  {Rydh}},\ }\href@noop {} {\bibfield  {journal} {\bibinfo  {journal} {Review
  of Scientific Instruments}\ }\textbf {\bibinfo {volume} {83}},\ \bibinfo
  {pages} {055107} (\bibinfo {year} {2012})}\BibitemShut {NoStop}%
\bibitem [{\citenamefont {Fortune}\ and\ \citenamefont
  {Hannahs}(2014)}]{fortune_top-loading_2014}%
  \BibitemOpen
  \bibfield  {author} {\bibinfo {author} {\bibfnamefont {N.~A.}\ \bibnamefont
  {Fortune}}\ and\ \bibinfo {author} {\bibfnamefont {S.~T.}\ \bibnamefont
  {Hannahs}},\ }\href@noop {} {\bibfield  {journal} {\bibinfo  {journal}
  {Journal of Physics: Conference Series}\ }\textbf {\bibinfo {volume} {568}},\
  \bibinfo {pages} {032008} (\bibinfo {year} {2014})}\BibitemShut {NoStop}%
\bibitem [{\citenamefont {Fortune}\ \emph {et~al.}(2000)\citenamefont
  {Fortune}, \citenamefont {Gossett}, \citenamefont {Peabody}, \citenamefont
  {Lehe}, \citenamefont {Uji},\ and\ \citenamefont {Aoki}}]{fortune_high_2000}%
  \BibitemOpen
  \bibfield  {author} {\bibinfo {author} {\bibfnamefont {N.}~\bibnamefont
  {Fortune}}, \bibinfo {author} {\bibfnamefont {G.}~\bibnamefont {Gossett}},
  \bibinfo {author} {\bibfnamefont {L.}~\bibnamefont {Peabody}}, \bibinfo
  {author} {\bibfnamefont {K.}~\bibnamefont {Lehe}}, \bibinfo {author}
  {\bibfnamefont {S.}~\bibnamefont {Uji}},\ and\ \bibinfo {author}
  {\bibfnamefont {H.}~\bibnamefont {Aoki}},\ }\href@noop {} {\bibfield
  {journal} {\bibinfo  {journal} {Review of Scientific Instruments}\ }\textbf
  {\bibinfo {volume} {71}},\ \bibinfo {pages} {3825} (\bibinfo {year}
  {2000})}\BibitemShut {NoStop}%
\bibitem [{\citenamefont {Rosa}\ and\ \citenamefont
  {Fisk}(2021)}]{Rosa2021BulkAS}%
  \BibitemOpen
  \bibfield  {author} {\bibinfo {author} {\bibfnamefont {P.~F.~S.}\
  \bibnamefont {Rosa}}\ and\ \bibinfo {author} {\bibfnamefont {Z.}~\bibnamefont
  {Fisk}},\ }\bibfield  {title} {\bibinfo {title} {Bulk and surface properties
  of \text{SmB}\textsubscript{6}},\ }in\ \href@noop {} {\emph {\bibinfo
  {booktitle} {Rare-Earth Borides}}},\ \bibinfo {editor} {edited by\ \bibinfo
  {editor} {\bibfnamefont {D.~S.}\ \bibnamefont {Inosov}}}\ (\bibinfo
  {publisher} {Jenny Stanford Publishing},\ \bibinfo {address} {New York},\
  \bibinfo {year} {2021})\ Chap.~\bibinfo {chapter} {11}\BibitemShut {NoStop}%
\bibitem [{\citenamefont {Ikeda}\ \emph {et~al.}(1991)\citenamefont {Ikeda},
  \citenamefont {Dhar}, \citenamefont {Yoshizawa},\ and\ \citenamefont
  {Gschneidner}}]{ikeda_quenching_1991}%
  \BibitemOpen
  \bibfield  {author} {\bibinfo {author} {\bibfnamefont {K.}~\bibnamefont
  {Ikeda}}, \bibinfo {author} {\bibfnamefont {S.}~\bibnamefont {Dhar}},
  \bibinfo {author} {\bibfnamefont {M.}~\bibnamefont {Yoshizawa}},\ and\
  \bibinfo {author} {\bibfnamefont {K.}~\bibnamefont {Gschneidner}},\
  }\href@noop {} {\bibfield  {journal} {\bibinfo  {journal} {Journal of
  Magnetism and Magnetic Materials}\ }\textbf {\bibinfo {volume} {100}},\
  \bibinfo {pages} {292} (\bibinfo {year} {1991})}\BibitemShut {NoStop}%
\bibitem [{\citenamefont {Fuhrman}\ and\ \citenamefont
  {Nikolić}(2020)}]{fuhrman_magnetic_2020}%
  \BibitemOpen
  \bibfield  {author} {\bibinfo {author} {\bibfnamefont {W.~T.}\ \bibnamefont
  {Fuhrman}}\ and\ \bibinfo {author} {\bibfnamefont {P.}~\bibnamefont
  {Nikolić}},\ }\href@noop {} {\bibfield  {journal} {\bibinfo  {journal}
  {Physical Review B}\ }\textbf {\bibinfo {volume} {101}},\ \bibinfo {pages}
  {245118} (\bibinfo {year} {2020})}\BibitemShut {NoStop}%
\bibitem [{\citenamefont {Gopal}(1966)}]{gopal_esr_specific_1966}%
  \BibitemOpen
  \bibfield  {author} {\bibinfo {author} {\bibfnamefont {E.}~\bibnamefont
  {Gopal}},\ }\href@noop {} {\emph {\bibinfo {title} {Specific {Heats} at {Low}
  {Temperatures}}}}\ (\bibinfo  {publisher} {Springer},\ \bibinfo {year}
  {1966})\BibitemShut {NoStop}%
\bibitem [{\citenamefont {Flachbart}\ \emph {et~al.}(2002)\citenamefont
  {Flachbart}, \citenamefont {Gabáni}, \citenamefont {Gloos}, \citenamefont
  {Konovalova}, \citenamefont {Orendac}, \citenamefont {Paderno}, \citenamefont
  {Pavlik}, \citenamefont {Reiffers},\ and\ \citenamefont
  {Samuely}}]{FLACHBART_2002}%
  \BibitemOpen
  \bibfield  {author} {\bibinfo {author} {\bibfnamefont {K.}~\bibnamefont
  {Flachbart}}, \bibinfo {author} {\bibfnamefont {S.}~\bibnamefont {Gabáni}},
  \bibinfo {author} {\bibfnamefont {K.}~\bibnamefont {Gloos}}, \bibinfo
  {author} {\bibfnamefont {E.}~\bibnamefont {Konovalova}}, \bibinfo {author}
  {\bibfnamefont {M.}~\bibnamefont {Orendac}}, \bibinfo {author} {\bibfnamefont
  {Y.}~\bibnamefont {Paderno}}, \bibinfo {author} {\bibfnamefont
  {V.}~\bibnamefont {Pavlik}}, \bibinfo {author} {\bibfnamefont
  {M.}~\bibnamefont {Reiffers}},\ and\ \bibinfo {author} {\bibfnamefont
  {P.}~\bibnamefont {Samuely}},\ }\href
  {https://doi.org/https://doi.org/10.1016/S0921-4526(01)01127-9} {\bibfield
  {journal} {\bibinfo  {journal} {Physica B: Condensed Matter}\ }\textbf
  {\bibinfo {volume} {312-313}},\ \bibinfo {pages} {379} (\bibinfo {year}
  {2002})}\BibitemShut {NoStop}%
\bibitem [{\citenamefont {Gab{\'a}ni}\ \emph {et~al.}(2001)\citenamefont
  {Gab{\'a}ni}, \citenamefont {Flachbart}, \citenamefont {Konovalova},
  \citenamefont {Orend{\'a}c}, \citenamefont {Paderno}, \citenamefont
  {Pavl{\'i}k},\ and\ \citenamefont {Sebek}}]{Gabni_2001}%
  \BibitemOpen
  \bibfield  {author} {\bibinfo {author} {\bibfnamefont {S.}~\bibnamefont
  {Gab{\'a}ni}}, \bibinfo {author} {\bibfnamefont {K.}~\bibnamefont
  {Flachbart}}, \bibinfo {author} {\bibfnamefont {E.~S.}\ \bibnamefont
  {Konovalova}}, \bibinfo {author} {\bibfnamefont {M.}~\bibnamefont
  {Orend{\'a}c}}, \bibinfo {author} {\bibfnamefont {Y.~B.}\ \bibnamefont
  {Paderno}}, \bibinfo {author} {\bibfnamefont {V.}~\bibnamefont
  {Pavl{\'i}k}},\ and\ \bibinfo {author} {\bibfnamefont {J.}~\bibnamefont
  {Sebek}},\ }\href@noop {} {\bibfield  {journal} {\bibinfo  {journal} {Solid
  State Communications}\ }\textbf {\bibinfo {volume} {117}},\ \bibinfo {pages}
  {641} (\bibinfo {year} {2001})}\BibitemShut {NoStop}%
\bibitem [{\citenamefont {Flachbart}\ \emph {et~al.}(2006)\citenamefont
  {Flachbart}, \citenamefont {Gabáni}, \citenamefont {Neumaier}, \citenamefont
  {Paderno}, \citenamefont {Pavlík}, \citenamefont {Schuberth},\ and\
  \citenamefont {Shitsevalova}}]{flachbart_specific_2006}%
  \BibitemOpen
  \bibfield  {author} {\bibinfo {author} {\bibfnamefont {K.}~\bibnamefont
  {Flachbart}}, \bibinfo {author} {\bibfnamefont {S.}~\bibnamefont {Gabáni}},
  \bibinfo {author} {\bibfnamefont {K.}~\bibnamefont {Neumaier}}, \bibinfo
  {author} {\bibfnamefont {Y.}~\bibnamefont {Paderno}}, \bibinfo {author}
  {\bibfnamefont {V.}~\bibnamefont {Pavlík}}, \bibinfo {author} {\bibfnamefont
  {E.}~\bibnamefont {Schuberth}},\ and\ \bibinfo {author} {\bibfnamefont
  {N.}~\bibnamefont {Shitsevalova}},\ }\href@noop {} {\bibfield  {journal}
  {\bibinfo  {journal} {Physica B: Condensed Matter}\ ,\ \bibinfo {pages}
  {610}} (\bibinfo {year} {2006})}\BibitemShut {NoStop}%
\bibitem [{\citenamefont {Stewart}(2001)}]{stewart_non-fermi-liquid_2001}%
  \BibitemOpen
  \bibfield  {author} {\bibinfo {author} {\bibfnamefont {G.~R.}\ \bibnamefont
  {Stewart}},\ }\href {https://doi.org/10.1103/RevModPhys.73.797} {\bibfield
  {journal} {\bibinfo  {journal} {Reviews of Modern Physics}\ }\textbf
  {\bibinfo {volume} {73}},\ \bibinfo {pages} {797} (\bibinfo {year}
  {2001})}\BibitemShut {NoStop}%
\bibitem [{\citenamefont {Gheidi}\ \emph {et~al.}(2019)\citenamefont {Gheidi},
  \citenamefont {Akintola}, \citenamefont {Akella}, \citenamefont {Côté},
  \citenamefont {Dunsiger}, \citenamefont {Broholm}, \citenamefont {Fuhrman},
  \citenamefont {Saha}, \citenamefont {Paglione},\ and\ \citenamefont
  {Sonier}}]{gheidi_intrinsic_2019}%
  \BibitemOpen
  \bibfield  {author} {\bibinfo {author} {\bibfnamefont {S.}~\bibnamefont
  {Gheidi}}, \bibinfo {author} {\bibfnamefont {K.}~\bibnamefont {Akintola}},
  \bibinfo {author} {\bibfnamefont {K.}~\bibnamefont {Akella}}, \bibinfo
  {author} {\bibfnamefont {A.}~\bibnamefont {Côté}}, \bibinfo {author}
  {\bibfnamefont {S.}~\bibnamefont {Dunsiger}}, \bibinfo {author}
  {\bibfnamefont {C.}~\bibnamefont {Broholm}}, \bibinfo {author} {\bibfnamefont
  {W.}~\bibnamefont {Fuhrman}}, \bibinfo {author} {\bibfnamefont
  {S.}~\bibnamefont {Saha}}, \bibinfo {author} {\bibfnamefont {J.}~\bibnamefont
  {Paglione}},\ and\ \bibinfo {author} {\bibfnamefont {J.}~\bibnamefont
  {Sonier}},\ }\href@noop {} {\bibfield  {journal} {\bibinfo  {journal}
  {Physical Review Letters}\ }\textbf {\bibinfo {volume} {123}},\ \bibinfo
  {pages} {197203} (\bibinfo {year} {2019})}\BibitemShut {NoStop}%
\bibitem [{\citenamefont {Stewart}\ \emph {et~al.}(1984)\citenamefont
  {Stewart}, \citenamefont {Fisk}, \citenamefont {Willis},\ and\ \citenamefont
  {Smith}}]{stewart_possibility_1984}%
  \BibitemOpen
  \bibfield  {author} {\bibinfo {author} {\bibfnamefont {G.~R.}\ \bibnamefont
  {Stewart}}, \bibinfo {author} {\bibfnamefont {Z.}~\bibnamefont {Fisk}},
  \bibinfo {author} {\bibfnamefont {J.~O.}\ \bibnamefont {Willis}},\ and\
  \bibinfo {author} {\bibfnamefont {J.~L.}\ \bibnamefont {Smith}},\ }\href@noop
  {} {\bibfield  {journal} {\bibinfo  {journal} {Physical Review Letters}\
  }\textbf {\bibinfo {volume} {52}},\ \bibinfo {pages} {679} (\bibinfo {year}
  {1984})}\BibitemShut {NoStop}%
\bibitem [{\citenamefont {Brinkman}\ and\ \citenamefont
  {Engelsberg}(1968)}]{brinkman_spin-fluctuation_1968}%
  \BibitemOpen
  \bibfield  {author} {\bibinfo {author} {\bibfnamefont {W.}~\bibnamefont
  {Brinkman}}\ and\ \bibinfo {author} {\bibfnamefont {S.}~\bibnamefont
  {Engelsberg}},\ }\href@noop {} {\bibfield  {journal} {\bibinfo  {journal}
  {Physical Review}\ }\textbf {\bibinfo {volume} {169}},\ \bibinfo {pages}
  {417} (\bibinfo {year} {1968})}\BibitemShut {NoStop}%
\bibitem [{\citenamefont {Phelan}\ \emph {et~al.}(2014)\citenamefont {Phelan},
  \citenamefont {Koohpayeh}, \citenamefont {Cottingham}, \citenamefont
  {Freeland}, \citenamefont {Leiner}, \citenamefont {Broholm},\ and\
  \citenamefont {McQueen}}]{phelan_correlation_2014}%
  \BibitemOpen
  \bibfield  {author} {\bibinfo {author} {\bibfnamefont {W.}~\bibnamefont
  {Phelan}}, \bibinfo {author} {\bibfnamefont {S.}~\bibnamefont {Koohpayeh}},
  \bibinfo {author} {\bibfnamefont {P.}~\bibnamefont {Cottingham}}, \bibinfo
  {author} {\bibfnamefont {J.}~\bibnamefont {Freeland}}, \bibinfo {author}
  {\bibfnamefont {J.}~\bibnamefont {Leiner}}, \bibinfo {author} {\bibfnamefont
  {C.}~\bibnamefont {Broholm}},\ and\ \bibinfo {author} {\bibfnamefont
  {T.}~\bibnamefont {McQueen}},\ }\href@noop {} {\bibfield  {journal} {\bibinfo
   {journal} {Physical Review X}\ }\textbf {\bibinfo {volume} {4}},\ \bibinfo
  {pages} {031012} (\bibinfo {year} {2014})}\BibitemShut {NoStop}%
\bibitem [{\citenamefont {Orendáč}\ \emph {et~al.}(2017)\citenamefont
  {Orendáč}, \citenamefont {Gabáni}, \citenamefont {Pristáš},
  \citenamefont {Gažo}, \citenamefont {Diko}, \citenamefont {Farkašovský},
  \citenamefont {Levchenko}, \citenamefont {Shitsevalova},\ and\ \citenamefont
  {Flachbart}}]{orendac_isosbestic_2017}%
  \BibitemOpen
  \bibfield  {author} {\bibinfo {author} {\bibfnamefont {M.}~\bibnamefont
  {Orendáč}}, \bibinfo {author} {\bibfnamefont {S.}~\bibnamefont {Gabáni}},
  \bibinfo {author} {\bibfnamefont {G.}~\bibnamefont {Pristáš}}, \bibinfo
  {author} {\bibfnamefont {E.}~\bibnamefont {Gažo}}, \bibinfo {author}
  {\bibfnamefont {P.}~\bibnamefont {Diko}}, \bibinfo {author} {\bibfnamefont
  {P.}~\bibnamefont {Farkašovský}}, \bibinfo {author} {\bibfnamefont
  {A.}~\bibnamefont {Levchenko}}, \bibinfo {author} {\bibfnamefont
  {N.}~\bibnamefont {Shitsevalova}},\ and\ \bibinfo {author} {\bibfnamefont
  {K.}~\bibnamefont {Flachbart}},\ }\href@noop {} {\bibfield  {journal}
  {\bibinfo  {journal} {Physical Review B}\ }\textbf {\bibinfo {volume} {96}},\
  \bibinfo {pages} {115101} (\bibinfo {year} {2017})}\BibitemShut {NoStop}%
\bibitem [{\citenamefont {Hertel}\ \emph {et~al.}(1980)\citenamefont {Hertel},
  \citenamefont {Appel},\ and\ \citenamefont {Fay}}]{hertel_effect_1980}%
  \BibitemOpen
  \bibfield  {author} {\bibinfo {author} {\bibfnamefont {P.}~\bibnamefont
  {Hertel}}, \bibinfo {author} {\bibfnamefont {J.}~\bibnamefont {Appel}},\ and\
  \bibinfo {author} {\bibfnamefont {D.}~\bibnamefont {Fay}},\ }\href@noop {}
  {\bibfield  {journal} {\bibinfo  {journal} {Physical Review B}\ }\textbf
  {\bibinfo {volume} {22}},\ \bibinfo {pages} {534} (\bibinfo {year}
  {1980})}\BibitemShut {NoStop}%
\bibitem [{\citenamefont {Béal-Monod}\ and\ \citenamefont
  {Daniel}(1983)}]{beal-monod_field_1983}%
  \BibitemOpen
  \bibfield  {author} {\bibinfo {author} {\bibfnamefont {M.~T.}\ \bibnamefont
  {Béal-Monod}}\ and\ \bibinfo {author} {\bibfnamefont {E.}~\bibnamefont
  {Daniel}},\ }\href@noop {} {\bibfield  {journal} {\bibinfo  {journal}
  {Physical Review B}\ }\textbf {\bibinfo {volume} {27}},\ \bibinfo {pages}
  {4467} (\bibinfo {year} {1983})}\BibitemShut {NoStop}%
\bibitem [{\citenamefont {Caldwell}\ \emph {et~al.}(2007)\citenamefont
  {Caldwell}, \citenamefont {Reyes}, \citenamefont {Moulton}, \citenamefont
  {Kuhns}, \citenamefont {Hoch}, \citenamefont {Schlottmann},\ and\
  \citenamefont {Fisk}}]{caldwell_high-field_2007}%
  \BibitemOpen
  \bibfield  {author} {\bibinfo {author} {\bibfnamefont {T.}~\bibnamefont
  {Caldwell}}, \bibinfo {author} {\bibfnamefont {A.~P.}\ \bibnamefont {Reyes}},
  \bibinfo {author} {\bibfnamefont {W.~G.}\ \bibnamefont {Moulton}}, \bibinfo
  {author} {\bibfnamefont {P.~L.}\ \bibnamefont {Kuhns}}, \bibinfo {author}
  {\bibfnamefont {M.~J.~R.}\ \bibnamefont {Hoch}}, \bibinfo {author}
  {\bibfnamefont {P.}~\bibnamefont {Schlottmann}},\ and\ \bibinfo {author}
  {\bibfnamefont {Z.}~\bibnamefont {Fisk}},\ }\href@noop {} {\bibfield
  {journal} {\bibinfo  {journal} {Physical Review B}\ }\textbf {\bibinfo
  {volume} {75}},\ \bibinfo {pages} {075106} (\bibinfo {year}
  {2007})}\BibitemShut {NoStop}%
\bibitem [{\citenamefont
  {Dingle}(1952{\natexlab{a}})}]{dingle_magnetic_1952-1}%
  \BibitemOpen
  \bibfield  {author} {\bibinfo {author} {\bibfnamefont {R.~B.}\ \bibnamefont
  {Dingle}},\ }\href@noop {} {\bibfield  {journal} {\bibinfo  {journal}
  {Proceedings of the Royal Society of London. Series A}\ }\textbf {\bibinfo
  {volume} {211}},\ \bibinfo {pages} {500} (\bibinfo {year}
  {1952}{\natexlab{a}})}\BibitemShut {NoStop}%
\bibitem [{\citenamefont
  {Dingle}(1952{\natexlab{b}})}]{dingle_magnetic_1952-2}%
  \BibitemOpen
  \bibfield  {author} {\bibinfo {author} {\bibfnamefont {R.}~\bibnamefont
  {Dingle}},\ }\href@noop {} {\bibfield  {journal} {\bibinfo  {journal}
  {Proceedings of the Royal Society of London Series A}\ }\textbf {\bibinfo
  {volume} {211}},\ \bibinfo {pages} {517} (\bibinfo {year}
  {1952}{\natexlab{b}})}\BibitemShut {NoStop}%
\bibitem [{\citenamefont {VanderPlas}(2018)}]{vanderplas_understanding_2018}%
  \BibitemOpen
  \bibfield  {author} {\bibinfo {author} {\bibfnamefont {J.~T.}\ \bibnamefont
  {VanderPlas}},\ }\href {https://doi.org/10.3847/1538-4365/aab766} {\bibfield
  {journal} {\bibinfo  {journal} {The Astrophysical Journal Supplement Series}\
  }\textbf {\bibinfo {volume} {236}},\ \bibinfo {pages} {16} (\bibinfo {year}
  {2018})}\BibitemShut {NoStop}%
\bibitem [{\citenamefont {Ishizawa}\ \emph {et~al.}(1977)\citenamefont
  {Ishizawa}, \citenamefont {Tanaka}, \citenamefont {Bannai},\ and\
  \citenamefont {Kawai}}]{ishizawa_haas-van_1977}%
  \BibitemOpen
  \bibfield  {author} {\bibinfo {author} {\bibfnamefont {Y.}~\bibnamefont
  {Ishizawa}}, \bibinfo {author} {\bibfnamefont {T.}~\bibnamefont {Tanaka}},
  \bibinfo {author} {\bibfnamefont {E.}~\bibnamefont {Bannai}},\ and\ \bibinfo
  {author} {\bibfnamefont {S.}~\bibnamefont {Kawai}},\ }\href@noop {}
  {\bibfield  {journal} {\bibinfo  {journal} {Journal of the Physical Society
  of Japan}\ }\textbf {\bibinfo {volume} {42}},\ \bibinfo {pages} {112}
  (\bibinfo {year} {1977})}\BibitemShut {NoStop}%
\bibitem [{\citenamefont {Reifenberger}\ and\ \citenamefont
  {Stark}(1979)}]{reifenberger_1979}%
  \BibitemOpen
  \bibfield  {author} {\bibinfo {author} {\bibfnamefont {R.}~\bibnamefont
  {Reifenberger}}\ and\ \bibinfo {author} {\bibfnamefont {R.~W.}\ \bibnamefont
  {Stark}},\ }\href {https://doi.org/10.1088/0305-4608/9/7/011} {\bibfield
  {journal} {\bibinfo  {journal} {Journal of Physics F: Metal Physics}\
  }\textbf {\bibinfo {volume} {9}},\ \bibinfo {pages} {1307} (\bibinfo {year}
  {1979})}\BibitemShut {NoStop}%
\bibitem [{\citenamefont {Larson}\ and\ \citenamefont
  {Gordon}(1967)}]{larson_low-field_1967}%
  \BibitemOpen
  \bibfield  {author} {\bibinfo {author} {\bibfnamefont {C.~O.}\ \bibnamefont
  {Larson}}\ and\ \bibinfo {author} {\bibfnamefont {W.~L.}\ \bibnamefont
  {Gordon}},\ }\href {https://doi.org/10.1103/PhysRev.156.703} {\bibfield
  {journal} {\bibinfo  {journal} {Physical Review}\ }\textbf {\bibinfo {volume}
  {156}},\ \bibinfo {pages} {703} (\bibinfo {year} {1967})}\BibitemShut
  {NoStop}%
\bibitem [{\citenamefont {Zhang}\ \emph {et~al.}(2020)\citenamefont {Zhang},
  \citenamefont {Singh}, \citenamefont {Lane}, \citenamefont {Kidd},
  \citenamefont {Zhang}, \citenamefont {Barbiellini}, \citenamefont
  {Markiewicz}, \citenamefont {Bansil},\ and\ \citenamefont
  {Sun}}]{zhang_understanding_2020}%
  \BibitemOpen
  \bibfield  {author} {\bibinfo {author} {\bibfnamefont {R.}~\bibnamefont
  {Zhang}}, \bibinfo {author} {\bibfnamefont {B.}~\bibnamefont {Singh}},
  \bibinfo {author} {\bibfnamefont {C.}~\bibnamefont {Lane}}, \bibinfo {author}
  {\bibfnamefont {J.}~\bibnamefont {Kidd}}, \bibinfo {author} {\bibfnamefont
  {Y.}~\bibnamefont {Zhang}}, \bibinfo {author} {\bibfnamefont
  {B.}~\bibnamefont {Barbiellini}}, \bibinfo {author} {\bibfnamefont {R.~S.}\
  \bibnamefont {Markiewicz}}, \bibinfo {author} {\bibfnamefont
  {A.}~\bibnamefont {Bansil}},\ and\ \bibinfo {author} {\bibfnamefont
  {J.}~\bibnamefont {Sun}},\ }\href {http://arxiv.org/abs/2003.11052}
  {\bibfield  {journal} {\bibinfo  {journal} {arXiv:2003.11052 [cond-mat]}\ }
  (\bibinfo {year} {2020})}\BibitemShut {NoStop}%
\bibitem [{\citenamefont {Harrison}(2018)}]{harrison_highly_2018a}%
  \BibitemOpen
  \bibfield  {author} {\bibinfo {author} {\bibfnamefont {N.}~\bibnamefont
  {Harrison}},\ }\href@noop {} {\bibfield  {journal} {\bibinfo  {journal}
  {Phys. Rev. Lett.}\ }\textbf {\bibinfo {volume} {121}},\ \bibinfo {pages}
  {026602} (\bibinfo {year} {2018})}\BibitemShut {NoStop}%
\bibitem [{\citenamefont {Shen}\ and\ \citenamefont
  {Fu}(2018)}]{shen_quantum_2018}%
  \BibitemOpen
  \bibfield  {author} {\bibinfo {author} {\bibfnamefont {H.}~\bibnamefont
  {Shen}}\ and\ \bibinfo {author} {\bibfnamefont {L.}~\bibnamefont {Fu}},\
  }\href@noop {} {\bibfield  {journal} {\bibinfo  {journal} {Physical Review
  Letters}\ }\textbf {\bibinfo {volume} {121}},\ \bibinfo {pages} {026403}
  (\bibinfo {year} {2018})}\BibitemShut {NoStop}%
\bibitem [{\citenamefont {Xiang}\ \emph {et~al.}(2021)\citenamefont {Xiang},
  \citenamefont {Chen}, \citenamefont {Chen}, \citenamefont {Tinsman},
  \citenamefont {Sato}, \citenamefont {Asaba}, \citenamefont {Lu},
  \citenamefont {Kasahara}, \citenamefont {Jaime}, \citenamefont {Balakirev},
  \citenamefont {Iga}, \citenamefont {Matsuda}, \citenamefont {Singleton},\
  and\ \citenamefont {Li}}]{Xiang_YbB12}%
  \BibitemOpen
  \bibfield  {author} {\bibinfo {author} {\bibfnamefont {Z.}~\bibnamefont
  {Xiang}}, \bibinfo {author} {\bibfnamefont {L.}~\bibnamefont {Chen}},
  \bibinfo {author} {\bibfnamefont {K.-W.}\ \bibnamefont {Chen}}, \bibinfo
  {author} {\bibfnamefont {C.}~\bibnamefont {Tinsman}}, \bibinfo {author}
  {\bibfnamefont {Y.}~\bibnamefont {Sato}}, \bibinfo {author} {\bibfnamefont
  {T.}~\bibnamefont {Asaba}}, \bibinfo {author} {\bibfnamefont
  {H.}~\bibnamefont {Lu}}, \bibinfo {author} {\bibfnamefont {Y.}~\bibnamefont
  {Kasahara}}, \bibinfo {author} {\bibfnamefont {M.}~\bibnamefont {Jaime}},
  \bibinfo {author} {\bibfnamefont {F.}~\bibnamefont {Balakirev}}, \bibinfo
  {author} {\bibfnamefont {F.}~\bibnamefont {Iga}}, \bibinfo {author}
  {\bibfnamefont {Y.}~\bibnamefont {Matsuda}}, \bibinfo {author} {\bibfnamefont
  {J.}~\bibnamefont {Singleton}},\ and\ \bibinfo {author} {\bibfnamefont
  {L.}~\bibnamefont {Li}},\ }\href {https://doi.org/10.1038/s41567-021-01216-0}
  {\bibfield  {journal} {\bibinfo  {journal} {Nature Physics}\ }\textbf
  {\bibinfo {volume} {17}},\ \bibinfo {pages} {788} (\bibinfo {year}
  {2021})}\BibitemShut {NoStop}%
\end{thebibliography}%


\providecommand{\noopsort}[1]{}\providecommand{\singleletter}[1]{#1}%
\begin{thebibliography}{10}
\expandafter\ifx\csname natexlab\endcsname\relax\def\natexlab#1{#1}\fi
\expandafter\ifx\csname bibnamefont\endcsname\relax
  \def\bibnamefont#1{#1}\fi
\expandafter\ifx\csname bibfnamefont\endcsname\relax
  \def\bibfnamefont#1{#1}\fi
\expandafter\ifx\csname citenamefont\endcsname\relax
  \def\citenamefont#1{#1}\fi
\expandafter\ifx\csname url\endcsname\relax
  \def\url#1{\texttt{#1}}\fi
\expandafter\ifx\csname urlprefix\endcsname\relax\def\urlprefix{URL }\fi
\providecommand{\bibinfo}[2]{#2}
\providecommand{\eprint}[2][]{\url{#2}}

\bibitem[{\citenamefont{Orendáč et~al.}(2017)\citenamefont{Orendáč,
  Gabáni, Pristáš, Gažo, Diko, Farkašovský, Levchenko, Shitsevalova, and
  Flachbart}}]{orendac_isosbestic_2017}
\bibinfo{author}{\bibfnamefont{M.}~\bibnamefont{Orendáč}},
  \bibinfo{author}{\bibfnamefont{S.}~\bibnamefont{Gabáni}},
  \bibinfo{author}{\bibfnamefont{G.}~\bibnamefont{Pristáš}},
  \bibinfo{author}{\bibfnamefont{E.}~\bibnamefont{Gažo}},
  \bibinfo{author}{\bibfnamefont{P.}~\bibnamefont{Diko}},
  \bibinfo{author}{\bibfnamefont{P.}~\bibnamefont{Farkašovský}},
  \bibinfo{author}{\bibfnamefont{A.}~\bibnamefont{Levchenko}},
  \bibinfo{author}{\bibfnamefont{N.}~\bibnamefont{Shitsevalova}},
  \bibnamefont{and}
  \bibinfo{author}{\bibfnamefont{K.}~\bibnamefont{Flachbart}},
  \bibinfo{journal}{Physical Review B} \textbf{\bibinfo{volume}{96}},
  \bibinfo{pages}{115101} (\bibinfo{year}{2017}), \bibinfo{note}{publisher:
  American Physical Society},
  \urlprefix\url{https://link.aps.org/doi/10.1103/PhysRevB.96.115101}.

\bibitem[{\citenamefont{Mandrus et~al.}(2001)\citenamefont{Mandrus, Sales, and
  Jin}}]{mandrus_localized_2001}
\bibinfo{author}{\bibfnamefont{D.}~\bibnamefont{Mandrus}},
  \bibinfo{author}{\bibfnamefont{B.~C.} \bibnamefont{Sales}}, \bibnamefont{and}
  \bibinfo{author}{\bibfnamefont{R.}~\bibnamefont{Jin}},
  \bibinfo{journal}{Physical Review B} \textbf{\bibinfo{volume}{64}}
  (\bibinfo{year}{2001}), ISSN \bibinfo{issn}{0163-1829},
  \bibinfo{note}{publisher: American Physical Society (APS)},
  \urlprefix\url{https://dx.doi.org/10.1103/physrevb.64.012302}.

\bibitem[{\citenamefont{Trounov et~al.}(1993)\citenamefont{Trounov, Malyshev,
  Chernyshov, Korsukova, Gurin, Aslanov, and
  Chernyshev}}]{trounov_temperature_1993}
\bibinfo{author}{\bibfnamefont{V.~A.} \bibnamefont{Trounov}},
  \bibinfo{author}{\bibfnamefont{A.~L.} \bibnamefont{Malyshev}},
  \bibinfo{author}{\bibfnamefont{D.~Y.} \bibnamefont{Chernyshov}},
  \bibinfo{author}{\bibfnamefont{M.~M.} \bibnamefont{Korsukova}},
  \bibinfo{author}{\bibfnamefont{V.~N.} \bibnamefont{Gurin}},
  \bibinfo{author}{\bibfnamefont{L.~A.} \bibnamefont{Aslanov}},
  \bibnamefont{and} \bibinfo{author}{\bibfnamefont{V.~V.}
  \bibnamefont{Chernyshev}}, \bibinfo{journal}{Journal of Physics: Condensed
  Matter} \textbf{\bibinfo{volume}{5}}, \bibinfo{pages}{2479}
  (\bibinfo{year}{1993}), ISSN \bibinfo{issn}{0953-8984},
  \bibinfo{note}{publisher: IOP Publishing},
  \urlprefix\url{https://doi.org/10.1088/0953-8984/5/16/007}.

\bibitem[{\citenamefont{Phelan et~al.}(2014)\citenamefont{Phelan, Koohpayeh,
  Cottingham, Freeland, Leiner, Broholm, and
  McQueen}}]{phelan_correlation_2014}
\bibinfo{author}{\bibfnamefont{W.}~\bibnamefont{Phelan}},
  \bibinfo{author}{\bibfnamefont{S.}~\bibnamefont{Koohpayeh}},
  \bibinfo{author}{\bibfnamefont{P.}~\bibnamefont{Cottingham}},
  \bibinfo{author}{\bibfnamefont{J.}~\bibnamefont{Freeland}},
  \bibinfo{author}{\bibfnamefont{J.}~\bibnamefont{Leiner}},
  \bibinfo{author}{\bibfnamefont{C.}~\bibnamefont{Broholm}}, \bibnamefont{and}
  \bibinfo{author}{\bibfnamefont{T.}~\bibnamefont{McQueen}},
  \bibinfo{journal}{Physical Review X} \textbf{\bibinfo{volume}{4}},
  \bibinfo{pages}{031012} (\bibinfo{year}{2014}), \bibinfo{note}{publisher:
  American Physical Society},
  \urlprefix\url{https://link.aps.org/doi/10.1103/PhysRevX.4.031012}.

\bibitem[{\citenamefont{Schoenberg}(1984)}]{Schoenberg1984}
\bibinfo{author}{\bibfnamefont{D.}~\bibnamefont{Schoenberg}},
  \emph{\bibinfo{title}{Magnetic Oscillations in Metals}}
  (\bibinfo{publisher}{Cambridge University Press}, \bibinfo{year}{1984}).

\bibitem[{\citenamefont{VanderPlas}(2018)}]{vanderplas_understanding_2018}
\bibinfo{author}{\bibfnamefont{J.~T.} \bibnamefont{VanderPlas}},
  \bibinfo{journal}{The Astrophysical Journal Supplement Series}
  \textbf{\bibinfo{volume}{236}}, \bibinfo{pages}{16} (\bibinfo{year}{2018}),
  ISSN \bibinfo{issn}{1538-4365},
  \urlprefix\url{https://iopscience.iop.org/article/10.3847/1538-4365/aab766}.

\bibitem[{\citenamefont{Suzuki et~al.}(1985)\citenamefont{Suzuki, Goto,
  Fujimura, Kunii, Suzuki, and Kasuya}}]{suzuki_1985}
\bibinfo{author}{\bibfnamefont{T.}~\bibnamefont{Suzuki}},
  \bibinfo{author}{\bibfnamefont{T.}~\bibnamefont{Goto}},
  \bibinfo{author}{\bibfnamefont{T.}~\bibnamefont{Fujimura}},
  \bibinfo{author}{\bibfnamefont{S.}~\bibnamefont{Kunii}},
  \bibinfo{author}{\bibfnamefont{T.}~\bibnamefont{Suzuki}}, \bibnamefont{and}
  \bibinfo{author}{\bibfnamefont{T.}~\bibnamefont{Kasuya}},
  \bibinfo{journal}{Journal of Magnetism and Magnetic Materials}
  \textbf{\bibinfo{volume}{52}}, \bibinfo{pages}{261} (\bibinfo{year}{1985}),
  ISSN \bibinfo{issn}{0304-8853},
  \urlprefix\url{https://www.sciencedirect.com/science/article/pii/0304885385902744}.

\bibitem[{\citenamefont{Ishizawa et~al.}(1977)\citenamefont{Ishizawa, Tanaka,
  Bannai, and Kawai}}]{Ishizawa}
\bibinfo{author}{\bibfnamefont{Y.}~\bibnamefont{Ishizawa}},
  \bibinfo{author}{\bibfnamefont{T.}~\bibnamefont{Tanaka}},
  \bibinfo{author}{\bibfnamefont{E.}~\bibnamefont{Bannai}}, \bibnamefont{and}
  \bibinfo{author}{\bibfnamefont{S.}~\bibnamefont{Kawai}},
  \bibinfo{journal}{Journal of the Physical Society of Japan}
  \textbf{\bibinfo{volume}{42}}, \bibinfo{pages}{112} (\bibinfo{year}{1977}),
  ISSN \bibinfo{issn}{0031-9015},
  \urlprefix\url{https://doi.org/10.1143/JPSJ.42.112}.

\bibitem[{\citenamefont{Thalmeier et~al.}(1987)\citenamefont{Thalmeier,
  Lemmens, Ewert, Lenz, and Winzer}}]{Thalmeier}
\bibinfo{author}{\bibfnamefont{P.}~\bibnamefont{Thalmeier}},
  \bibinfo{author}{\bibfnamefont{P.}~\bibnamefont{Lemmens}},
  \bibinfo{author}{\bibfnamefont{S.}~\bibnamefont{Ewert}},
  \bibinfo{author}{\bibfnamefont{D.}~\bibnamefont{Lenz}}, \bibnamefont{and}
  \bibinfo{author}{\bibfnamefont{K.}~\bibnamefont{Winzer}},
  \bibinfo{journal}{Europhysics Letters (EPL)} \textbf{\bibinfo{volume}{4}},
  \bibinfo{pages}{1177} (\bibinfo{year}{1987}), ISSN \bibinfo{issn}{0295-5075
  1286-4854}, \urlprefix\url{http://dx.doi.org/10.1209/0295-5075/4/10/016}.

\bibitem[{\citenamefont{Onuki et~al.}(1989)\citenamefont{Onuki, Umezawa, Kwok,
  Crabtree, Nishihara, Yamazaki, Omi, and Komatsubara}}]{Onuki}
\bibinfo{author}{\bibfnamefont{Y.}~\bibnamefont{Onuki}},
  \bibinfo{author}{\bibfnamefont{A.}~\bibnamefont{Umezawa}},
  \bibinfo{author}{\bibfnamefont{W.~K.} \bibnamefont{Kwok}},
  \bibinfo{author}{\bibfnamefont{G.~W.} \bibnamefont{Crabtree}},
  \bibinfo{author}{\bibfnamefont{M.}~\bibnamefont{Nishihara}},
  \bibinfo{author}{\bibfnamefont{T.}~\bibnamefont{Yamazaki}},
  \bibinfo{author}{\bibfnamefont{T.}~\bibnamefont{Omi}}, \bibnamefont{and}
  \bibinfo{author}{\bibfnamefont{T.}~\bibnamefont{Komatsubara}},
  \bibinfo{journal}{Phys Rev B Condens Matter} \textbf{\bibinfo{volume}{40}},
  \bibinfo{pages}{11195} (\bibinfo{year}{1989}), ISSN \bibinfo{issn}{0163-1829
  (Print) 0163-1829}.

\end{thebibliography}

\end{document}


\title{Supplemental Information for: Observation of Quantum Oscillations in The Low Temperature Specific Heat of SmB\textsubscript{6}}

\author{P. G.  LaBarre}
\email{pglabarre@gmail.com}
\affiliation{Department of Physics, University of California at Santa Cruz, Santa Cruz, CA, USA}

\author{A. Rydh}
\affiliation{Department of Physics, Stockholm University, Stockholm, Sweden}

\author{J. Palmer-Fortune}
\affiliation{Department of Physics, Smith College, Northampton MA 01063}

\author{J. A. Frothingham}
\affiliation{Department of Physics, Smith College, Northampton MA 01063}


\author{S. T. Hannahs} 
\affiliation{National High Magnetic Field Laboratory, Florida State University, Tallahassee, FL 32310-3706, USA}



\author{A. P. Ramirez }
\email{apr@ucsc.edu}
\affiliation{Department of Physics, University of California at Santa Cruz, Santa Cruz, CA, USA}

\author{N. Fortune}
\email{nfortune@smith.edu}
\affiliation{Department of Physics, Smith College, Northampton MA 01063}

\date{\today}

\maketitle

\section{Non-oscillatory contributions to the specific heat of S\lowercase{m}B$_6$}

 
In order to model the 'high' temperature behavior ($T > T^*$) we followed a similar approach to Orendáč \textit{et al.} \cite{orendac_isosbestic_2017}, shown in Eq. (2) 
\begin{equation}
    C/T = \gamma + C_{E} 
\end{equation}
where $C_E$ is the specific heat from the Einstein model. They first infer a lattice contribution to the specific heat from corresponding measurements of isomorphous $LaB_6$, using a model by Mandrus \textit{et al.} \cite{mandrus_localized_2001} for $LaB_6$ which treats the La ions as independent Einstein oscillators embedded in a boron framework treated as a Debeye solid. 
 
In that model, the Debeye $\Theta_D$ and Einstein $\Theta_E$ oscillator temperatures are determined by x-ray diffraction to be $\Theta_D = 1160 \text{ K}$ and $\Theta_E = 141 \text{ K}$. Inclusion of the Einstein term greatly improves the fit to the specific data by accounting for a prominent shoulder in the data due to local vibrations of La ions within a rigid 3D Boron ``cage."  For the specific heat calculation, 1 mol of La ions are treated as Einstein oscillators and 6 mol of B ions are treated as a Debye solid.
  
For a monatomic cubic crystal, $U_{iso}(T)$ can be solved exactly within the Debye approximation: 
  \begin{equation}\label{eq:U_Debeye}
  U_{iso} = \left[ \frac{3h^2 T}{4 \pi^2 m k_b {\Theta_D}^2}\right] \left[ \Phi(\Theta_D) + \frac{1}{4}\frac{\Theta_D}{T}\right] 
  \end{equation}
where
  \begin{equation}
      \label{eq:Phi_Debeye}
      \Phi(x) = \frac{1}{x}\int_0^x dy \frac{y}{e^y - 1}
  \end{equation}
For $LaB_6$, $U_{iso} = 0.0040(1) \text{ \AA}^2 = 0.00004 \text{ nm}^2$ yields a value $\Theta_D = 1160 \text{ K}$. Correspondingly, for an Einstein oscillator, the mean square displacement amplitude is given by 
  \begin{equation}
      \label{eq:U_Einstein}
      U_{iso} = \frac{h^2}{8 \pi^2 m k_B \Theta_E}\coth{\left(\frac{\Theta_E}{2 T}\right)}
  \end{equation}
and for $La$, $U_{iso} = 0.00537(2) {\text{ \AA}}^2 = 5.37 \cdot  10^{-23} m^2$, corresponding to an Einstein temperature $\Theta_E = 141 \text{ K}$. 
 
Based on the corresponding room temperature x-ray diffraction data for $SmB_6$ \cite{trounov_temperature_1993}, we calculate $\Theta_E = 119 \text{ K}$ 
consistent with Trounov's \textit{et al.} \cite{trounov_temperature_1993} estimate of  Einstein temperature $\Theta_E = 120 \text{ K}$. The Einstein model for the contribution of La or Sm atom 'rattling around' in a B 'cage' is given by 
\begin{equation}
    \label{C_Einstein}
    C_E = 3 s R \left( \frac{\Theta_E}{T}\right)^2
    \frac{ \exp{ \left(  \Theta_E / T \right) }  }
    { \left[ \exp{ \left(\Theta_E / T \right)}- 1\right]^2}.
\end{equation}
Applying this model for $T > 15 K$ we find good fits to our experimental data with $\gamma_0 = 35.7 \frac{mJ}{moleK^2}$ and $\Theta_E = 106K$, shown by the dashed cyan line in Fig. 1 above 15 K.

\begin{figure}[ht!]
\includegraphics[width=0.7\columnwidth]{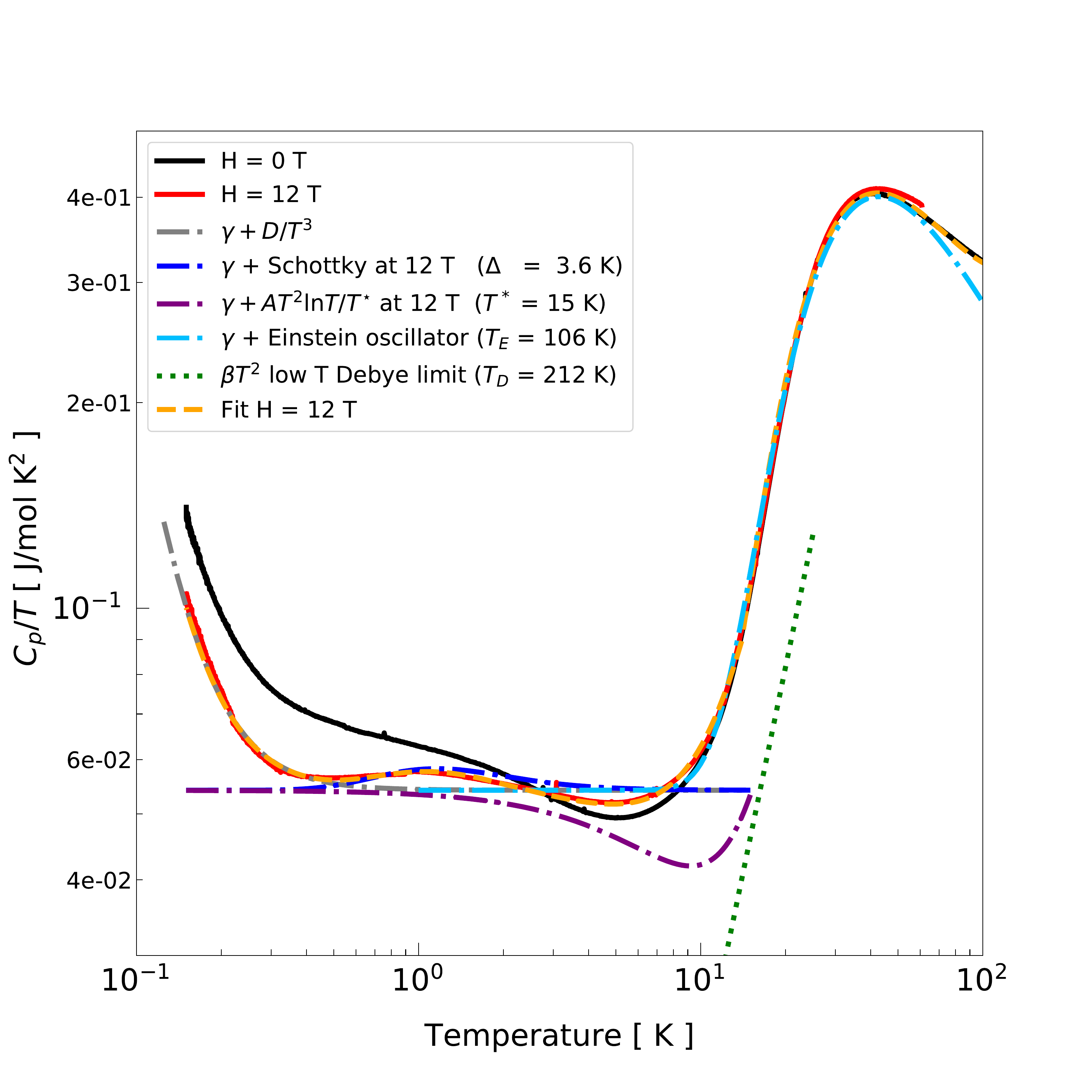}
\caption{\label{fig:fullfit} Temperature dependence of the specific heat at 0 T and 12 T for a 0.430 $\mu$g flux-grown SmB\textsubscript{6} sample, along with a representative theoretical fit to the temperature dependence at 12 T. The results are consistent with the presence of spin fluctuations with a characteristic temperature \emph{\(T^{\ast}\)} = 15 K.}
\end{figure}

As discussed in the main text at both  0T and 12 T our data are well described at low $T$ ($T < 15 K$) by: 
\begin{equation}
\label{eq:model_fit}
\begin{split}
C &  =C_{el} + C_{\KI} + \beta_{D}T^3 +  DT^{-2} \\
C_{el} & = \gamma_0 T\ \left[\left(m^{\star }/{m}\right)+A T^2 \ln{\left( {T}/{T^{\star }} \right)} \right] 
\end{split}
\end{equation}
where $\gamma_0$ in $C_{el}$ is the ``bare'' electronic coefficient of the specific heat expected from band structure, \( m^{\ast}/{m}=\gamma\left(H\right)/{\gamma_0}\) is the many-body effective mass enhancement above the band mass $m$, $A$ is a coupling constant dependent on the strength of the exchange interaction between Fermi-liquid quasiparticles and mass-enhancing excitations, $D/T^2$ is an empirically determined fitting term for the lowest temperature behavior.

We find in good agreement with previous works \cite{phelan_correlation_2014, orendac_isosbestic_2017}, $\gamma(H = 12T) = 54.1 \frac{mJ}{mole K^2}$, $T_D = 212 K$, $A = 0.287 \frac{mJ}{moleK^4}$, and $D = 0.156 \frac{mJK}{mole}$

\section{Lifshitz-Kosevich (L-K) Fitting Procedure}

\begin{figure}[ht!]
    \centering
    \includegraphics[width=0.8\columnwidth]{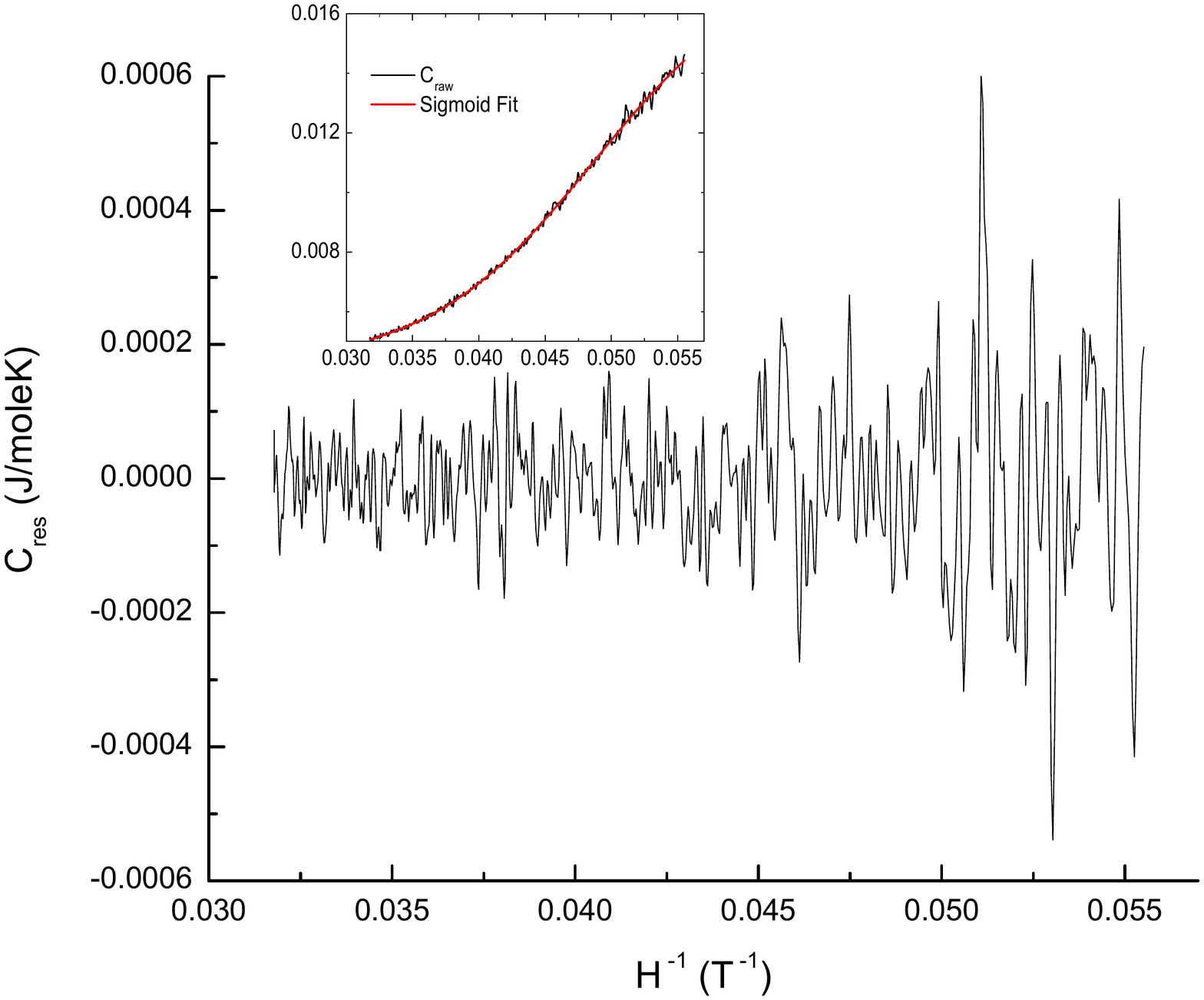}
    \caption{$C_{res}$ v $H^{-1}$ at $T = 0.58 K$ for $H = 18-31$ T. Inset Sigmoid fit to the raw specific heat data, the black curve is $C_{raw}$ and the red curve is the sigmoid fit used in the background subtraction.}
    \label{cres}
\end{figure}

In order to detect oscillations at the $< 0.01\% \gamma_0$ level a background subtraction was made using a sigmoid distribution, $y =  \frac{A_1-A_2}{1+e^{(x-x_0)/dx}} +A_2$, where $A_1$, $A_2$, $x_0$ and $dx$ are fitting parameters from $H = 18-31$ T  as is shown in the inset of Fig. \ref{cres} for increasing field at $T = 0.58 K$.  The sigmoid function was chosen due to its non-oscillatory nature over any field range and a shape that is similar to the raw data above 14 T.  By contrast, fitting the background signal to higher order polynomials can lead to artificial oscillatory residuals at the $0.01\% \gamma_0$ level near field regions where different powers of the polynomials become dominant.  Fig. \ref{cres} shows the residual specific heat for one of the field sweeps for $H = 18-31$ T at $T = 0.58 K$.

As is outlined by D. Shoenberg \cite{Schoenberg1984} the theoretical oscillatory specific heat derived via L-K theory in the limit of $T \rightarrow 0$ is 
\begin{equation}
    \Delta C_{osc} = -\frac{1}{T}\frac{m_e}{m^*}\left(\frac{m_e}{e\pi\hbar^2}\right)^{3/2}\left(2\mu_B\right)^{5/2}\frac{V}{p^{5/2}\pi^2(A^{''})^{1/2}}H^{5/2}cos\left[2\pi p\left(\frac{F}{H}+\phi \right)\right]
    \label{CzeroT}
\end{equation}
where $m_e$ is the electron mass, $m^*$ is the quasi particle effective mass, $\hbar$ is Plank’s constant, $\mu_B$ is the Bohr magneton, $V$ is the sample volume, $H$ is the applied external magnetic field, $F$ is the frequency of oscillation, $A^{''}$ is a dimensionless measure of the curvature of the Fermi surface and $p$ is an integer value that corresponds to the harmonic of the fundamental frequency ($p = 1$is the fundamental frequency).  The amplitude of the oscillatory thermodynamic potential, and thus the oscillatory specific heat, given by Eq. \ref{CzeroT} is reduced by the introduction of a finite scattering time due to i) impurity scattering, ii) phase smearing due to finite temperature effects, and iii) an effective mass dependent amplitude correction factor due to spin scattering. As temperature is increased, the specific heat reduces in oscillation amplitude and a restructuring of the oscillation envelope in certain field ranges.  At finite temperature, phase smearing alters Eq. \ref{CzeroT} to yield
\begin{equation}
    \Delta C_{osc} = -\frac{1}{T}\frac{m_e}{m^*}z^2f^{''}(z)\left(\frac{m_e}{e\pi\hbar^2}\right)^{3/2}\left(2\mu_B\right)^{5/2}\frac{V}{p^{5/2}\pi^2(A^{''})^{1/2}}H^{5/2}cos\left[2\pi p\left(\frac{F}{H}+\phi \right)\right]
    \label{Cphase}
\end{equation}
with $z=2\pi^2 p \frac{k_B T}{\beta^* H}$ and $f(z)=\frac{z}{sinh(z)}$ where $\beta^*=  \frac{e\hbar}{m^*}$.  Shown in Fig. 3(a) of the main text is the effect of phase smearing, $z^2 f^{''} (z)$, on the oscillatory waveform and overall envelope of oscillations. This introduces a node in the specific heat centered at a field that is governed by temperature and quasiparticle effective mass.  As is shown in Fig. 3 of the main text, for effective masses of 0.13, 3, 4, and 5 $m_e$ the specific heat node arises at a temperature of $T = 0.58 K$ in our field range of interest.  For effective masses ranging from 3-6 $m_e$ one expects a growth or reduction of the amplitude of magnetoquantum oscillations as the field passes through the node.  Introducing electron scattering results in an exponential damping of the oscillatory magnitude where Eq. \ref{Cphase} is multiplied by $R_D=e^{-p((2\pi^2 k_B)/\beta)T_D/H}$ where $T_D$ is the Dingle temperature, a sample dependent parameter. This damping effect is a controlling factor for setting the scale of possible oscillatory phenomena and necessitates low temperatures and large magnetic fields for oscillations to be large enough to be observed.  Here it is worth noting that the effective mass, $m^*$, enters into the temperature damping terms (phase smearing) but the bare electron mass, $m_e$, enters into the field damping terms (Dingle factor)\cite{Schoenberg1984}.  This complication, however, is commonly ignored and the effective mass is entered into the Dingle factor.  To properly compare our data to published literature, one needs to determine which fitting methods were used, since the calculated value of $T_{D,calc}$ for a given value of $R_D$ will differ from the true value $T_{D,true}$ by a factor of $m^*/m_e$ . Lastly, the introduction of spin scattering results in a constant reduction across all temperatures and fields of $R_s= cos\left(1/2 p\pi g m^*/m_e \right)$ where $g$ is the Landé $g$-factor.  One then finds the final finite temperature oscillatory specific heat including phase smearing, electron scattering, and spin splitting, to be that shown in Eq. \ref{Ctot} where $C_0 =  \frac{V}{\pi^2} \left(\frac{m_e}{2\pi \hbar^2 } \right)^{3/2} (2\mu_B )^{5/2}$ in SI units.
\begin{equation}
    \Delta C_{osc} = -\frac{1}{T}\frac{m_e}{m^*}z^2f^{''}(z) \frac{H^{5/2}}{(A^{''})^{1/2}}C_0 R_D R_s cos\left[2\pi p\left(\frac{F}{H}+\phi \right)\right]
    \label{Ctot}
\end{equation}
For a cylindrical Fermi surface (FS), the cross-sectional area is $A = \pi k_c^2$.  Then $A^{''}=2\pi$ and the magnitude of the specific heat oscillations is enhanced by $(A^{''})^{-1/2} \sim 0.4$. For a cubic FS, the cross-sectional area is $A=k_c^2$ such that the specific heat oscillation magnitude is enhanced by $(A^{''})^{-1/2} \sim 0.7$.  This assumes, however, that the FS spans the entirety of the BZ.  SmB$_6$ has a simple cubic lattice with $a=4.133 \textup{~\AA} $ where the area of the first BZ is $A_{BZ}=\left(\frac{2\pi}{a} \right)^2=2.31 \textup{~\AA}^{-2}$.  Scaling $A^{''}$ based on the ratio of the extremal FS area, calculated from the measured oscillation frequency using Onsager’s relation, $\Delta 1/H=\frac{2\pi e}{\hbar}  \frac{1}{A_e}$, to that of the first BZ, we have for a frequency of 695T,  $A_e⁄A_{BZ} =  0.06634/2.31 = 0.027$, which for a cylindrical FS enhances the oscillation amplitude by $\left(2\pi * 0.027\right)^{-1/2}=2.43$ and for a cubic FS enhances the oscillation amplitude by $\left(2*0.027\right)^{-1/2}=4.3$.

\section{Lomb-Scargle NDFT}

Traditional methods for studying quantum oscillatory phenomena in condensed matter utilize the standard ‘time’ series analysis of discrete Fourier transforms (DFT) or discrete fast Fourier transforms (DFFT).  For analyzing magnetoquantum oscillations, however, this procedure is only mathematically valid if the data are sampled uniformly in $1/H$. Since the data are obtained at regular time intervals as the magnet current is ramped, the analysis needs to compensate for the lack of periodicity in $1/H$.  This discrepancy is usually ignored when the oscillatory component is large and the field spacing between points is small. As the signal to noise ratio decreases, however, the non-uniformity of sampling can dramatically affect the observability of  MQOs.

Accurate determination of the power spectrum for low signal to noise non uniformly sampled data requires the use of non-uniform discrete Fourier transforms (NDFT) of which there are several methods.  One of the most useful for detecting non uniformly sampled low signal to noise sinusoidal periodicity is the Lomb-Scargle method \cite{vanderplas_understanding_2018}.  The Lomb-Scargle approach is beneficial over classical non uniform periodogram methods in that the noise distribution at each individual frequency is chi-square distributed under the null hypothesis where the periodogram results from Gaussian noise and the result is equivalent to a periodogram derived from a least squares analysis \cite{vanderplas_understanding_2018}.  
\begin{figure}[ht!]
    \centering
    \includegraphics[width=0.7\columnwidth]{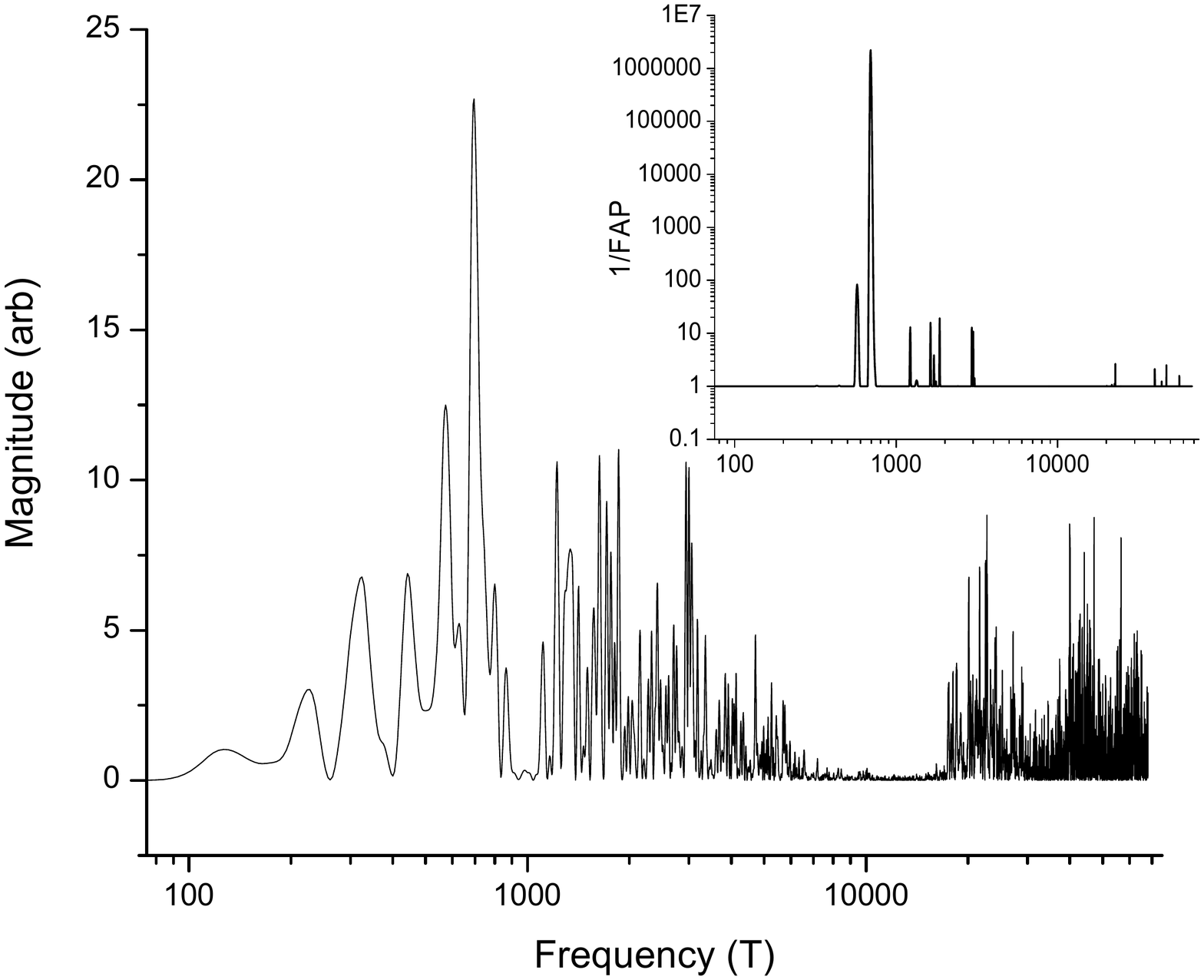}
    \caption{Unweighted frequency spectrum of Fig. \ref{cres}. Inset inverse false alarm probability used produce the weighted frequency spectrum shown in Fig. \ref{NDFT}.}
    \label{LSP}
\end{figure}
One important metric given by the Lomb-Scargle method is the false alarm probability (FAP).  The FAP is the probability that the resulting spectrum is composed of Gaussian noise and so a FAP = 1 indicates data at the specified frequency is best represented by Gaussian noise.  The closer the FAP is to zero, the higher the likelihood that the data at a specified frequency is a true frequency peak such that weighting the spectrum (Fig. \ref{LSP}) by the inverse of the FAP (Fig. \ref{LSP} inset) gives the best representation of the true frequency peaks, shown in Fig. \ref{NDFT}.  
\begin{figure}[ht!]
    \centering
    \includegraphics[width=0.7\columnwidth]{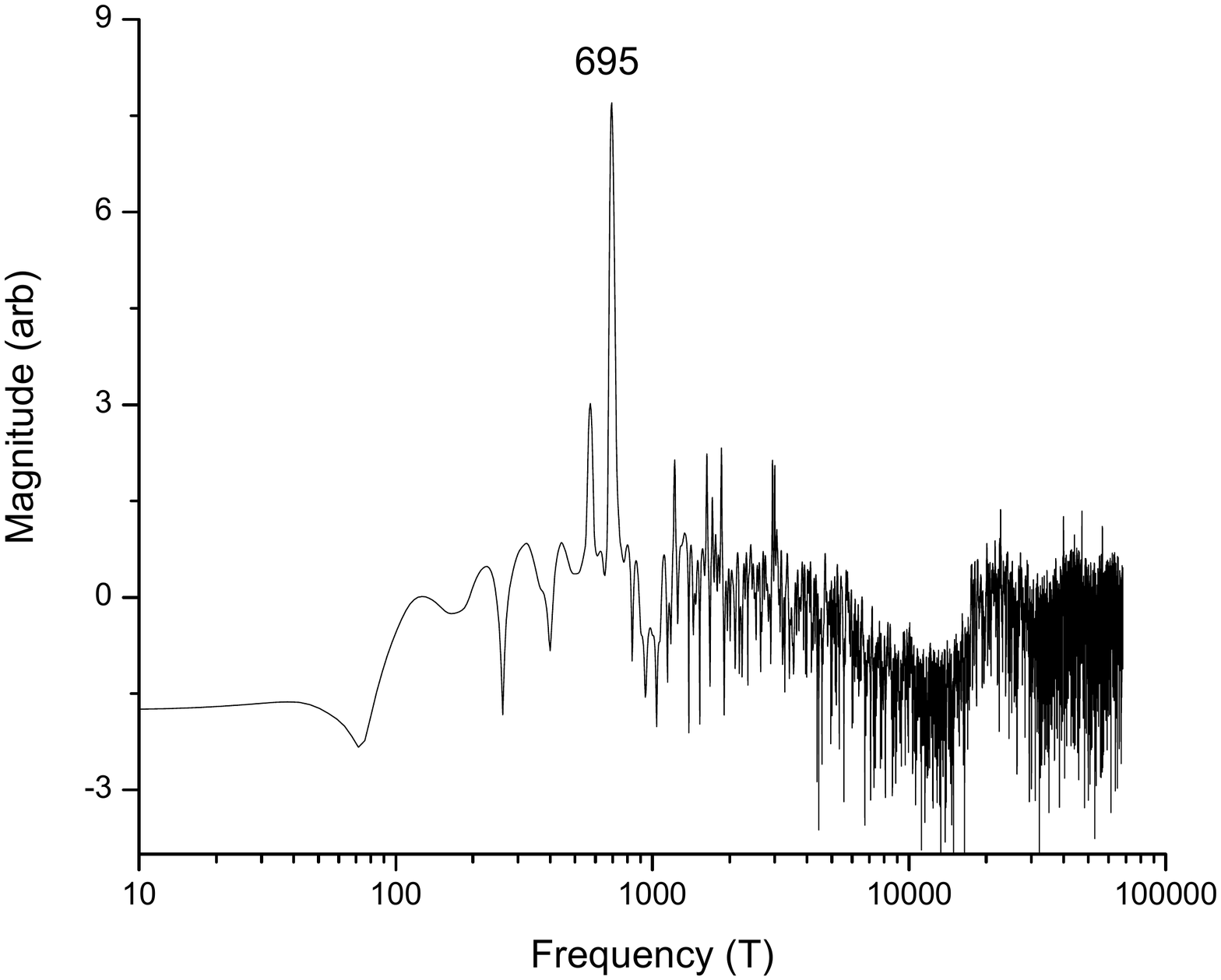}
    \caption{Logarithm of the weighted frequency spectrum generated from the residual heat capacity of SmB$_6$ at $T = 0.58 K$ with $H \vert\vert \Phi = 0$.}
    \label{NDFT}
\end{figure}

All data were prepared for frequency spectrum analysis in the same manner.  Utilizing the open source MATLAB code for Lomb-Scargle NUDFT made available by Jacob VanderPlas \cite{vanderplas_understanding_2018}, periodograms were generated with respect to crystallographic alignment with the applied external field for selected field sweeps. An oversampling factor of ten and a peak frequency of four times that of the average Nyquist frequency were used to make sure all possible oscillation frequencies were probed.

\section{L\lowercase{a}B$_6$}

In order to verify our methodology and application of L-K theory to classify the MQO’s in rare earth hexaborides we measured, from $H = 0-12$ T and $T = 1- 0.1 K$, $C$ of LaB$_6$, an isostructural metallic system that has shown large amplitude de-Haas van Alphen (dHvA) oscillations.  Measurements were made using a 12 T Blufors dilution refrigerator using custom-made membrane calorimeters at Stockholm University.  These calorimeters can accurately measure the specific heat of extremely small samples $(10 \mu m \times  30 \mu m \times 30 \mu m)$ and give the best experimentally achievable signal to noise for low temperature ac-calorimetry.  

\begin{figure}[ht!]
    \centering
    \includegraphics[width=0.7\columnwidth]{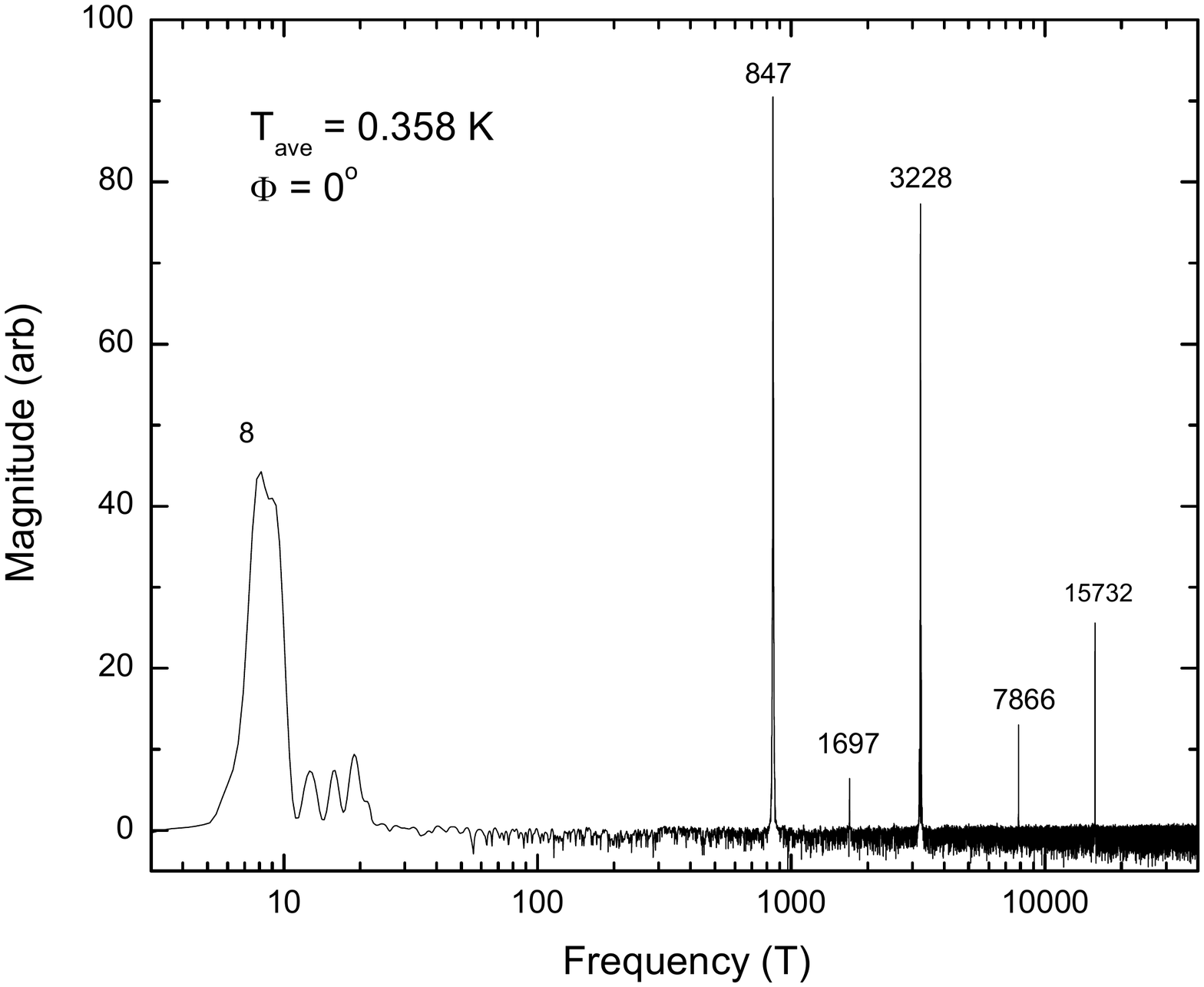}
    \caption{Peak magnitude vs. oscillation frequency at $T = 0.358K$ from $H = 2-12$ T for a LaB$_6$ sample $(10 \times 30 \times 70 \mu m^3)$.}
    \label{LaB6NDFT}
\end{figure}

Applying the Lomb-Scargle frequency analysis to the lowest noise measurements of LaB$_6$ at $T = 0.358 K$ we find the oscillation spectrum shown in Fig. \ref{LaB6NDFT}.  Here we find, for $\Phi = 0^o$, frequency peaks at $F = $8, 847, 1697, 3228, 7866, and 15732 T which correspond to the $\rho$, $\epsilon$, $2*\epsilon$, $\gamma$, $\alpha$, and $2*\alpha$ Fermi pockets respectively.  

\begin{figure}[ht!]
    \centering
    \includegraphics[width=0.7\columnwidth]{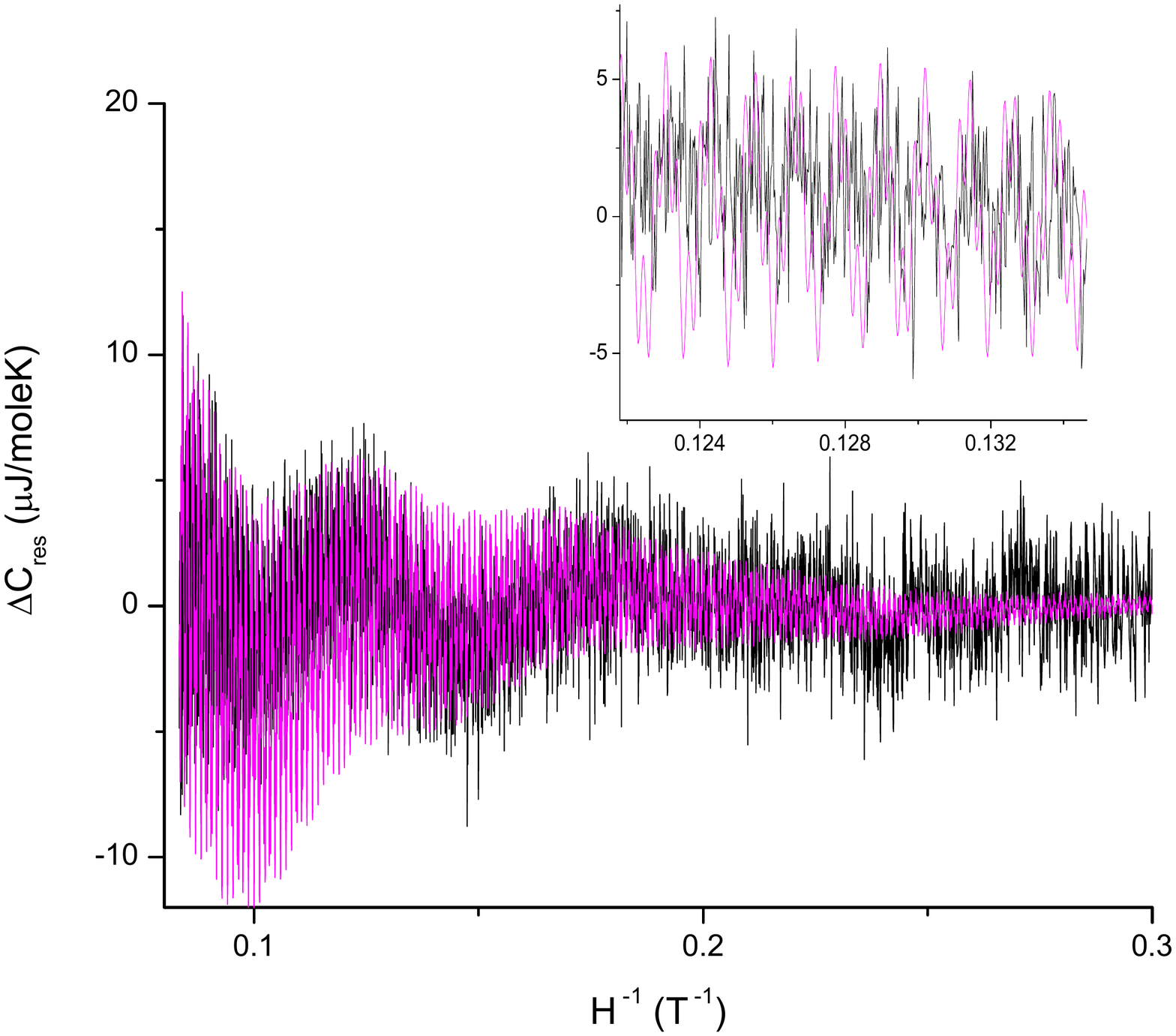}
    \caption{Residual $C$ vs $H^{-1}$ of LaB$_6$ at $T = 0.358 K$, the black curve is the raw data and the pink curve is the theoretical L-K fit of $\Delta C_{osc}$ using effective masses of $0.066 m_e$ for $\rho$ and $0.65 m_e$ for the $\epsilon$, $\gamma$, and $\alpha$ Fermi Pockets}
    \label{LaB6fit}
\end{figure}

These frequency values are in good agreement with values in the published literature \cite{suzuki_1985, Ishizawa, Thalmeier, Onuki}.  In Fig. \ref{LaB6fit} the black curve is the residual $C$ of LaB$_6$ resulting from the appropriate background subtraction and the pink curve is the theoretical L-K fit.  Using published values of effective mass for the $\rho$ and $\epsilon$ pockets for all frequency peaks $m_{eff} = 0.066- 0.65 m_e$ respectively and a Dingle temperature of $T_{D,true} = 0.6 K$ we find great agreement with the theoretical QO’s at $T = 0.358 K$.  The agreement of the frequency spectrum and resulting theoretical QO’s modeling via L-K theory for LaB$_6$ validates the use of this analysis for examining the residual specific heat of SmB$_6$.

\bibliographystyle{apsrev}
\bibliography{SmB6supp}